\documentclass[final,onefignum,onetabnum]{siamart190516}

\usepackage{tikz}
\usetikzlibrary{shapes,arrows, trees}
\usepackage{subcaption}
\usepackage{caption}

\usepackage{lipsum}
\usepackage[ddmmyyyy]{datetime}
\usepackage{amsfonts}
\usepackage{graphicx}
\usepackage{epstopdf}
\usepackage{algorithmic}
\ifpdf
  \DeclareGraphicsExtensions{,.pdf,.png,.jpg}
\else
  \DeclareGraphicsExtensions{}
\fi

\newcommand{\compresslist}{%
	\setlength{\itemsep}{0pt}%
	\setlength{\parskip}{1pt}%
	\setlength{\parsep}{0pt}%
}

\ifpdf
\hypersetup{
  pdftitle={SParSH-AMG: A library for hybrid CPU-GPU algebraic multigrid and preconditioned iterative methods},
  pdfauthor={Sashikumaar Ganesan, Manan Shah}
}
\fi

\headers{SParSH-AMG: Hybrid CPU-GPU algebraic multigrid Solver}{S. Ganesan, and M. Shah}

\title{SParSH-AMG: A library for hybrid CPU-GPU algebraic multigrid and preconditioned iterative methods\thanks{Submitted to the editors \today.
\funding{This work was partially supported by SERB, DST under contract no.~SERB-454.}
}}

\author{Sashikumaar Ganesan\thanks{Department of Computational and Data Sciences, Indian Institute of Science, Bangalore, India 560012  
  (\email{sashi@iisc.ac.in}, \email{shahjayant@iisc.ac.in}  ) } \and Manan Shah\footnotemark[2] 
}

\usepackage{amsopn}

\usepackage{todonotes}

\begin{document}

\maketitle
 
\begin{abstract}
Hybrid CPU-GPU algorithms for Algebraic Multigrid methods (AMG) to efficiently utilize both CPU and GPU resources are presented. In particular, hybrid AMG framework focusing on minimal utilization of GPU memory with performance on par with GPU-only implementations is developed. The hybrid AMG framework can be tuned to operate at a significantly lower GPU-memory, consequently, enables to solve large algebraic systems. Combining the  hybrid AMG framework as a preconditioner with Krylov Subspace solvers like Conjugate Gradient, BiCG methods provides a solver stack to solve a large class of problems.  The performance of the proposed hybrid AMG framework is analysed for an array of  matrices with different properties and size. Further, the performance of CPU-GPU algorithms are compared with the GPU-only implementations to illustrate the significantly lower memory requirements.

\end{abstract}



\section{Introduction}
Numerical simulation of   physical processes like fluid flow, heat transfer, etc  involves solving a large  system of linear equations. These linear or linearised systems are often obtained by discretization of partial differential equations (PDEs) that describe the  physical process. Moreover,   algebraic systems obtained by discretization of PDEs by  finite difference or finite element or finite volume methods are sparse in nature. Although direct solvers are robust and accurate, iterative solvers are preferred, especially for the solution of large systems, due to the high computational and memory requirements associated with the direct solvers.

In general, a slow convergence due to the persistence of  smooth error modes is one of the challenges associate with the solution of linear systems by a classical iterative method, especially for systems with a large condition number. The classical iterative methods such as Jacobi and Successive Over Relaxation (SOR)  do not suppress smooth components of the error. In particular, highly oscillatory error components  are damped rapidly by these methods, whereas smooth error  modes are  continue to persist in the solution. Multigrid method alleviates  this challenge by damping the smooth error modes on a coarse  system obtained from a coarse mesh or by coarsening the large system. The choice of coarse meshes to construct coarse systems leads to   Geometric Multigrid (GMG) method, whereas  coarse systems obtained by coarsening the large system leads to Algebraic Multigrid method.
Moreover, the order of complexity in multigrid methods is linear, $\mathcal{O}(N)$, where $N$ is the system size, that is, a system with $N$ degrees of freedoms (unknowns). Therefore, multigrid method is often the method of choice to solve such large sparse system of algebraic equations.

Algebraic multigrid method, unlike GMG, does not require access to mesh and other details of the physical problem. Hence, AMG can be used as a black-box solver or a preconditioner to other iterative methods. AMG involves construction of hierarchy of coarse matrices which are smaller in size than the original system and represent smooth modes of the error.  Classical coarsening approach proposed by Ruge and St\"uben~\cite{ruge1987algebraic} is one of the first coarsening strategy that selects subsets of fine level degrees of freedoms (DOFs) as coarse level DOFs. 
Many such heuristic based approaches such as Cleary-Luby-Jones-Plassman (CLJP) coarsening \cite{MO2007174}, Becks algorithm \cite{beck1999algebraic} have been proposed in the literature, see~\cite{stuben2001review} for an overview. In these approaches, local averaging of DOFs is performed to represent its value on the coarser level. Nevertheless, anisotropic problems require coarsening in particular directions and it becomes challenging with heuristic based approaches. Gandham $et. al.$~\cite{gandham2014gpu} have proposed an aggregation based coarsening strategies to overcome these limitations. Further, Notay~\cite{notay2010aggregation} has proposed an aggregation based coarsening approach using a heavy edge matching (HEM) algorithm. 
Two variants of HEM are proposed in this paper.


Solving sparse linear systems with AMG is a compute intensive operation. Such computations are often need  to be parallelized to reduce overall solution time. Modern day workstations and supercomputers are equipped with multicore CPUs and accelerators like GPU to facilitate faster computations. Utilization of such hybrid architectures for compute intensive application like AMG method requires redesigning of existing implementations. Hypre~\cite{falgout2002hypre} provides a scalable implementation of solvers for distributed environments. AMGX~\cite{naumov2015amgx} and BootCMatch~\cite{d2018bootcmatch} are GPU based AMG packages which can be used in single GPU or multi-GPU environments. These GPU-only solver libraries provide better performance and are optimised for specialized hardware. Nevertheless, CPU resources   are rich nowadays and are often underutilized in these GPU-only implementations. Moreover, GPU device memory is limited and it is one of the main limitations of GPU-only AMG implementations to solve large systems. In addition, data transfer latency is also a challenge when the system is large. These challenges necessitates to develop and implement algorithms that utilize both CPUs and GPUs resources efficiently and effectively. It is the key objective of this paper. 


GPU accelerators are often used as shared resources among multiple CPU cores on a workstation or on each multi-core node in supercomputers. Offloading compute intensive operations to the associated GPU by all CPU cores at a time  often results in shortfall of available GPU resources.
Moreover, it results in serialization of GPU calls and consequently the performance will be lost. Hence, a hybrid CPU-GPU approach is needed to efficiently utilize both CPUs and GPUs  and to improve the overall performance of the computation. It is the objective of this work  to design a hybrid CPU-GPU AMG implementation that utilizes both CPU and GPU resources optimally. The following contributions are made to achieve this objective:
\begin{itemize}
	\item Hybrid CPU-GPU AMG algorithms that require significantly minimal GPU resources (memory) without compromising the efficiency
	\item Enhancement of existing pairwise aggregation based coarsening strategy is formulated which utilize CSR matrix storage format for efficient formulation of inter-grid transfer operators
	\item Implementation of Krylov subspace solvers with hybrid CPU-GPU AMG as a preconditioner. 
\end{itemize}  

The paper is organized as follows: section \ref{sec_amg} describes components of AMG and improvement over existing pairwise coarsening approach. Parallel implementation of AMG with hybrid CPU-GPU approach and Krylov subspace solvers are described in Section \ref{sec_parallel}.   Numerical experiments that analyse the performance of the proposed parallel implementations are presented in Section~\ref{sec_experiments}. Finally, Section~\ref{sec_conc} ends with conclusion. 


\section{Algebraic Multigrid Method}
\label{sec_amg}
Let $A$ be a sparse matrix of size $N$, $u$ and $f$ be an unknown and a given column vectors of size $N$, respectively. Algebraic multigrid method  for the solution of a linear system
  \[
  Au=f
  \]
  consists the following three components:
\begin{itemize}
	
	\item \textbf{Smoothers:} Stationary iterative methods such as  Jacobi or Gauss-Siedal method is often used as a smoother to approximate hierarchy of algebraic systems. Non-stationary iterative methods such as Krylov subspace methods can also be used as a smoother provided it is efficient and has the property to damp highly oscillatory modes in the error.
	
	\item \textbf{Prolongation ($P$) and Restriction ($R$) Operators:}
	These operators transfer vectors between different finite-dimensional spaces. Restriction operator projects finite-dimensional function from a fine (high-dimensional) space  to a coarse space, whereas the prolongation performs the inverse operation. In AMG, these operators are linear mapping between the coarse and fine spaces. Suppose $N$  and $N_{c}$ are the dimensions of the fine  and coarse spaces respectively, then the prolongation   $P$ is $N$ $\times$ $N_{c}$ matrix, whereas the restriction $R$ is $P^{T}$.
	
	\item \textbf{Coarse Level Solver:} AMG requires a numerically-exact solution of the coarsest system, which is much smaller in size than the given system. Hence, a direct solver is often preferred to solve the coarse system. Moreover, it is enough to compute the LU factorization  of the coarse system only once and it can repeatedly be used  when the coarse system remains unchanged. Note that the numerical LU factorization needs to be computed whenever the values of the coarsest system matrix change. Nevertheless, it is enough to compute the symbolic LU factorization once. 
\end{itemize}{} 
\begin{figure}[htb!]
	\centering
	\tikzset{every picture/.style={line width=0.75pt}} 
	\begin{tikzpicture}[x=0.75pt,y=0.75pt,yscale=-1.5,xscale=1,scale=0.65]
	
	\draw    (216.5,31) -- (300.75,238.15) ;
	\draw [shift={(301.5,242)}, rotate = 247.87] [color={rgb, 255:red, 0; green, 0; blue, 0 }  ][line width=0.75]    (10.93,-3.29) .. controls (6.95,-1.4) and (3.31,-0.3) .. (0,0) .. controls (3.31,0.3) and (6.95,1.4) .. (10.93,3.29)   ;
	
	\draw    (301.5,242) -- (372.85,32.89) ;
	\draw [shift={(373.5,31)}, rotate = 468.84] [color={rgb, 255:red, 0; green, 0; blue, 0 }  ][line width=0.75]    (10.93,-3.29) .. controls (6.95,-1.4) and (3.31,-0.3) .. (0,0) .. controls (3.31,0.3) and (6.95,1.4) .. (10.93,3.29)   ;
	
	\draw    (151.5,242) -- (431.5,242) ;
	
	\draw    (114.5,31) -- (454.5,31) ;
		
	\draw (145,58) node [scale=1.0] [align=left] {Initial Guess u{\scriptsize 0}\\Presmooth};
	\draw (261,115) node [scale=1,rotate=-74] [align=left] {Restrict};
	\draw (175,144) node [scale=1]   {$ \begin{array}{l}
		r\ =\ f\ -\ A\widetilde{u}\\
		r\ =\ Au\ -\ A\widetilde{u}\\
		r\ =\ Ae\\
		r_{c} \ =\ P^{T} r\\
		A_{c} \ =\ P^{T} AP
		\end{array}$};
	\draw (300,260) node [scale=1]  {$A_{c} e_{c} \ =\ r_{
			c} \ ,e_{c} \ =\ A^{-1}_{c} r_{c}$};
	\draw (332,116) node [scale=1,rotate=-280] [align=left] {Prolongate};
	\draw (180,20) node [scale=1]  {$Au\ =\ f$};
	\draw (308,17) node [scale=1] [align=left] {Level 0};
	\draw (353,229) node [scale=1] [align=left] {Level 1};
	\draw (425,44) node [scale=1] [align=left] {Postsmooth};
	\draw (430,16) node [scale=1]  {$\widetilde{u} = \widetilde{u}\ +\ Pe_{c}$};
	\end{tikzpicture}
	\caption{Two Level Multigrid V-Cycle}
	\label{twolevelcycle}
\end{figure}
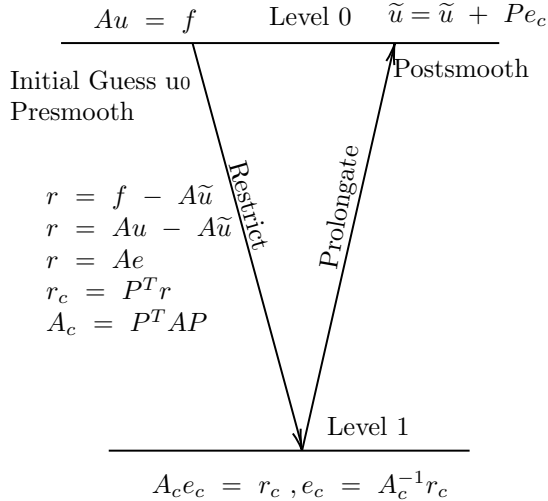{}
These components of the multigrid are combined in different order to form different multigrid cycles such as  V-cycle, W-cycle, F-cycle etc. Fig.~\ref{twolevelcycle} shows computations involved in two-level V-cycle. Algorithm \ref{algo_V} highlights the order in which these components are combined to form  a multigrid V-Cycle.
\begin{algorithm}
\caption{V-cycle$(k,f_{k},x_{k},A_{k})$}
\label{algo_V}
\label{alg:buildtree}
\begin{algorithmic}[1]
\STATE{\textbf{INPUT:} level $k$, rhs $f_{k}$, \textnormal{initial guess }$x_{k}$, \textnormal{Matrix }$A_{k}$}

\STATE{\textbf{OUTPUT:} updated solution $x_{k}$}

\STATE{	$x_{k}$   $\xleftarrow{}$ $S_{k}(f_{k},A_{k},x_{k})$ \null\hfill \textnormal{Presmoothing}\\}
\STATE{	$r_{k}$   $\xleftarrow{}$ $f_{k} - A_{k}x_{k}$       \null\hfill \textnormal{Compute Residual}\\} 
\STATE{$f_{k+1}$ $\xleftarrow{}$ $P^{T}_{k}r_{k}$  \null\hfill \textnormal{Restrict residual to coarser grid}\\}

	\IF{$k+1 = L$}
	\STATE{
		$x_{k+1}$ $\xleftarrow{}$ $x_{k}^{-1}f_{k+1}$ \null\hfill \textnormal{Solve the system exactly}\\
	}
	\ELSE
	\STATE{
		$x_{k+1}$ $\xleftarrow{}$ \textnormal{V-cycle}$(k+1,f_{k+1},0,A_{k+1})$ \null\hfill \textnormal{Recursion}\\
	}
    \ENDIF
    
    \STATE{$x_{k}$ $\xleftarrow{}$ $x_{k}$ + $P_{k}x_{k+1}$ \null\hfill \textnormal{Prolongate solution to finer grid}\\}
	
	\STATE{$x_{k}$  $\xleftarrow{}$ $S_{k}(f_{k},A_{k},x_{k})$ \null\hfill \textnormal{Post-smoothing}\\}
\RETURN $x_{k}$
\end{algorithmic}
\end{algorithm}
It can clearly be seen from the Algorithm~\ref{algo_V} that the multigrid method  solves the linear system by operating on hierarchy of coarse matrices. Unlike geometric multigrid method, where the hierarchy of coarse matrices are assembled from   uniformly refined meshes, the coarse matrices in AMG are constructed  by coarsening the given system matrix. In particular, we need to construct a transfer operator to perform coarsening of matrices and to transfer solutions across different level  of hierarchy. The transfer operators  play an important role in convergence characteristics of AMG. In particular, the coarse matrices obtained using the transfer operator should represent smooth components of the error, which are difficult to eliminate by smoothers. As mentioned in the introduction,   aggregation based AMG performs better than the classical approaches~\cite{gandham2014gpu}, especially when the coarsening needs to be performed in the directions of anisotropy. Heavy Edge Matching (HEM) coarsening approach \cite{notay2010aggregation} is one   of the aggregation based coarsening approaches, and two variants of HEM coarsening algorithms that take advantage of  CSR form of the system matrix to construct a transfer operator are presented in this section.



\subsection{Heavy Edge Matching Coarsening}
Assume that the matrix $A$ is given in Compressed Sparse Row (CSR) and is modeled as the adjacency matrix of a graph $G$, where each node $v_i,~0\le i<N,$ of the graph represents the DOF (the unknown) of the linear system. Further, let the coefficient $a_{ij},~0\le i,j<N,$ of the matrix $A$ be the edge-weight of the nodes $v_i$ and $v_j$. We apply graph coarsening strategies to construct a coarse graph $G_c$ with less nodes, which in turn represents a coarse matrix $A_c$ of the matrix $A$. 
Contrast to the classical coarsening approach, where a subset of nodes from the graph $G$ is selected to form a coarse graph $G_c$, aggregation based coarsening approach aggregates the nodes of the graph $G$ to form $G_c$. We propose two variants of HEM coarsening:
\begin{itemize}
    \item Node-based HEM algorithm
    \item Edge-based HEM algorithm
\end{itemize}
\subsubsection{Node-based HEM algorithm}  Initially all nodes of $G$ are marked as unmatched. Pairing of unmatched nodes is then performed, where each unmatched node is matched with its one of the unmatched neighbouring nodes that shares a highest absolute edge-weight. The matched pair of nodes are then aggregated and assigned a coarse node number. After that the aggregated pairs of nodes are marked as matched nodes. Suppose an unmatched node   does not find a pair, that is, does not have an unmatched neighbour, then the unmatched node is marked as matched but unaggregated and assigned a coarse node number to it. This differs from the previous HEM algorithms \cite{naumov2015amgx,notay2010aggregation}, where all unaggregated nodes are aggregated with it nearest nodes.  Since all unaggregated nodes are considered as coarser DOFs, we expect a better projections of vectors across the levels. In addition, matching is performed alternately from the fist and the last indices of the node to get a uniform coarsening. Such heuristics enable to maintain consistent coarsening ratios across all the levels. These coarsening steps are given in algorithm~\ref{HEM}.

\begin{figure}[tbh]
		\centering
		\begin{subfigure}{.65\textwidth}
			\centering
			\includegraphics[scale=0.25]{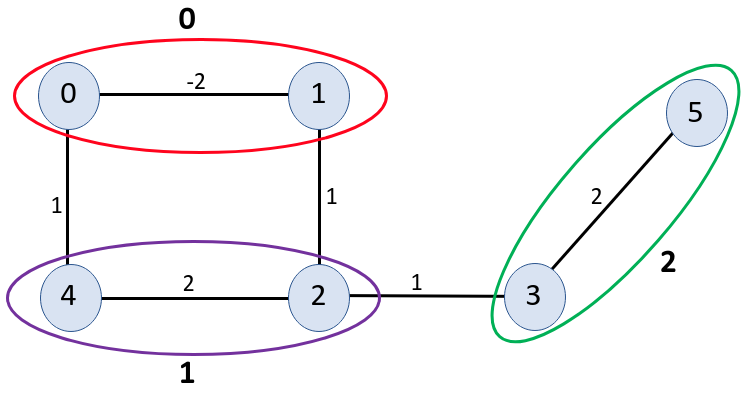}
			\caption{Graph for Matrix A}
			\label{fig:sub1}
		\end{subfigure}%
		\begin{subfigure}{.35\textwidth}
			\centering
			\includegraphics[scale=0.13]{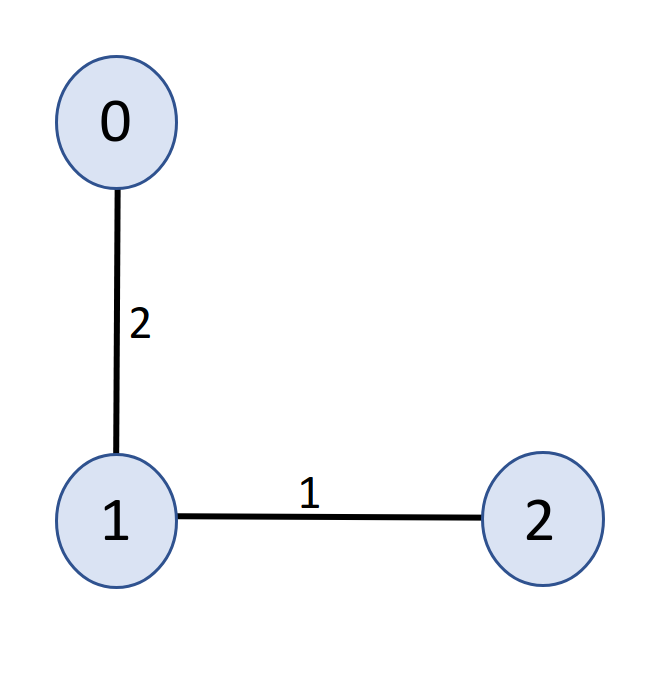}
			\caption{After Collapsing}
			\label{fig:sub2}
		\end{subfigure}
		\caption{Graph Coarsening by Node-based Heavy Edge Matching}
		\label{fig:test}
	\end{figure}

Simultaneously, the transfer operator, that is, the prolongation matrix $P$ is also constructed during the aggregation step in our algorithm. Suppose  nodes  $v_{i}$, $v_{j}$ in $G$ are aggregated and formed a coarse node $k$, then the $k^{th}$ column in $i^{th}$ and $j^{th}$ rows of the matrix $P$ will be non-zero. 
Though the structure of the matrix $P$ will remain same in all HEM approaches, the values of $P$ can be populated in different ways, see the remark at the end of this section. Here, the non-zero values of the matrix $P$  is populated with one.  Finally. the coarse matrix $A_{c}$ is obtained from the prolongation matrix by defining $A_{c} = P^TAP$.  In particular, the diagonal entries coarse level matrix $A_{c}$ are computed as follows:   $(A_{c})_{kk}$ = $a_{ij}$ + $a_{ji}$ + $a_{ii}$ + $a_{jj}$. For example, the graph of the matrix given  below  and its matching formed by algorithm~\ref{HEM} are shown in Fig~\ref{fig:sub1}. Further, Fig~\ref{fig:sub2} shows the coarse graph formed by the prolongation matrix $P$.
 \[
 A=
    \begin{bmatrix}
	4 & -2 & 0 & 0 & 1 & 0 \\
	-2 & 4 & 1 & 0 & 0 & 0 \\
	0 & 1 & 4 & 1 & 2 & 0 \\
	0 & 0 & 1 & 4 & 0 & 2 \\
	1 & 0 & 2 & 0 & 4 & 0 \\
	0 & 0 & 0 & 2 & 0 & 4 \\
	\end{bmatrix}
\quad 
P = 
	\begin{bmatrix}
	1 & 0 & 0 \\
	1 & 0 & 0 \\
	0 & 1 & 0 \\
	0 & 0 & 1 \\
	0 & 1 & 0 \\
	0 & 0 & 1 \\
	\end{bmatrix}
\quad
A_{c}  = 
	\begin{bmatrix}
	4 & 2 & 0 \\
	2 & 12 & 1 \\
	0 & 1 & 12 \\
	\end{bmatrix}
 \]

\begin{algorithm}
\caption{Node-based HEM algorithm}
\label{HEM}
\begin{algorithmic}[1]
\STATE{\textbf{INPUT:} $A$: $n$ $\times$ $n$ sparse matrix}

\STATE{\textbf{OUTPUT:}$P$ Matrix}

\STATE{	$I$ = \{0,....,$n-1$\} \null\hfill \textnormal{Set of unassigned vertices}\\}
\STATE{	$C$ = 0\null\hfill \textnormal{Initialize number of Coarse DOFs}\\} 
\FOR{i $\in$ I}
    \STATE{k = -1, match = -1\null\hfill \textnormal{Initialize Match}\\}
    \FOR{j such that j $\neq$ i , j $\in$ I and $a_{ij}$ $\neq$ $0$}
    
    \IF{max(k, abs($a_{ij}$)) =  abs($a_{ij}$) }
    \STATE{k = abs($a_{ij}$)\null\hfill \textnormal{Find the heaviest neighbour}\\}
    \STATE{match = j}
    
    \ENDIF
    \ENDFOR
    
    \IF{match $\neq$ -1}
	    \STATE{$P_{i,C}$ = 1\\}	    \STATE{$P_{match,C}$ = 1\\}
 	    \STATE{$C$ = $C$ + 1\\}	
 	    \STATE{$I = I - \{match\}$ \null\hfill \textnormal{Remove the matched vertex from unassigned list}\\}
 	\ELSE
	    \STATE{$P_{i,C}$ = 1\\}
	    \STATE{$C$ = $C$ + 1\\}
	\ENDIF
	\STATE{$I = I - \{i\}$ \null\hfill \textnormal{Remove the i vertex from unassigned list}}

\ENDFOR
\RETURN $P$ Matrix
\end{algorithmic}
\end{algorithm} 
Since unaggregated nodes are considered as coarse nodes, the coarsening ratio in proposed HEM algorithm is by a factor of slightly less than two. 
Though the proposed algorithm is inherently sequential, it avoids multiple visits to all nodes. \\

\noindent \textbf{Remark:}
 Node-based HEM algorithm is implemented as an unsmoothened aggregation approach, that is, the prolongation matrix $P$ is populate with one. Alternatively, a smooth vector $x$ obtained from the homogeneous system $Ax=0$ can also be used to populate the matrix $P$ and this approach is known as compatible weighted matching approach~\cite{bernaschi2020amg}.

\subsubsection{Edge-based HEM algorithm} 
A greedy approach is used in the proposed edge-based HEM coarsening algorithm  to pair the nodes, rather than the order of the node numbering. The remaining steps are same as in the node-based HEM algorithm given in the previous section. 
Initially,  a triple array containing the edge weight and its associated nodes for all edges is constructed. The array is then sorted in a decreasing order of edge-weights. After that each unmatched edge in the array is processed one by one and the associated nodes are aggregated provided that the nodes are unmatched. Finally, the aggregated nodes are marked as matched and assigned a coarse node number as in the node-based HEM algorithm. At the end, all unmatched nodes are left unaggregated and each unmatched node is assigned with a coarse node number.  The coarse matrices obtained from this algorithm remain invariant to numbering of the nodes.  
\begin{algorithm}
\caption{Edge-based HEM algorithm}
\label{HEM2}
\begin{algorithmic}[1]
\STATE{\textbf{INPUT:} $A$: $n$ $\times$ $n$ sparse matrix}
\STATE{\textbf{OUTPUT:}$P$ Matrix}

\STATE{	$I$ = \{0,....,$n-1$\} \null\hfill \textnormal{Set of unassigned vertices}\\}
\STATE{	$C$ = 0\null\hfill \textnormal{Initialize number of Coarse DOFs}\\}
\STATE{ $T$ = 0\null\hfill \textnormal{EdgeList: Stores tuples of (abs($a_{ij}$),i,j)}\\}
\FOR{i $\in$ I}
    \FOR{j such that j $\neq$ i and $a_{ij}$ $\neq$ $0$}
    \STATE{T $\xleftarrow{}$ T $\cup$ (abs($a_{ij}$),$i$,$j$)\null\hfill \textnormal{Add an edge to T set}\\}
    \ENDFOR
\ENDFOR
\STATE{Sort $T$ \null\hfill\textnormal{Sort T in lexicographical descending order}\\}
\WHILE{$T$ $\neq$ \{\}}
    \STATE{(abs($a_{ij}$),$i$,$j$) $\xleftarrow{}$ $T$ \null\hfill\textnormal{Pick the heaviest edge available in $T$ set}\\}
    \IF{ $i$ $\in$ $I$ and $j$ $\in$ $I$}
	    \STATE{$P_{i,C}$ = 1\null\hfill \textnormal{Assign DOFs a Coarse DOF number}\\}	    
	    \STATE{$P_{j,C}$ = 1\\}
 	    \STATE{$C$ = $C$ + 1\\}	
 	    \STATE{$I = I - \{i\} -\{j\}$  \null\hfill \textnormal{Remove the matched vertices from unassigned list}\\}
	\ENDIF
	\STATE{$T$ $\xleftarrow{}$ $T$ - (abs($a_{ij}$),$i$,$j$) \null\hfill \textnormal{Remove the edge from T set}\\}
\ENDWHILE

\FOR{$i$ $\in$ $I$}
    \STATE{$P_{i,C}$ = 1 \null\hfill \textnormal{Assign remaining DOFs Coarse DOF number}\\}
	\STATE{$C$ = $C$ + 1\\}
\ENDFOR
\RETURN $P$ Matrix
\end{algorithmic}
\end{algorithm} 
As in node-based coarsening,   the prolongation matrix $P$ is populated with one, that is, an unsmoothened aggregation approach is used. Algorithm \ref{HEM2} highlights the steps involved in the edge-based HEM algorithm. \\



\section{Parallel Implementations}
\label{sec_parallel}
The execution of AMG algorithms is split into two phases:
\begin{itemize}\compresslist
	\item \textbf{Setup Phase:} The setup phase involves all one-time operations such as memory allocations, construction of hierarchy of coarser matrices, computing LU factorization of  the coarsest  matrix  etc. This phase requires  access to the system matrix and consumes less than 10\% of total computing time  for a stationary problem. Therefore, the  setup phase is executed sequentially in order to avoid  communication overheads.  
	\item \textbf{Solve Phase:} The solve phase involves execution of prescribed multigrid cycles  such as  V- or W- or F-cycle. Moreover, it occupies major proportion of the total computing time.    
\end{itemize}
Furthermore, sparse linear solvers and preconditioners are I/O intensive applications in general. Their performance on current generation of processors is bandwidth bound. CPUs could not achieve higher FLOP rates on such applications due to low-bandwidth between CPU and DRAM. Nevertheless, the availability of  high-bandwidth    on accelerators such a Graphical Processing Units (GPUs)    make it suitable for compute intensive applications. In the following section, a few variants of hybrid CPU-GPU parallel implementations are proposed.

\subsection{AMG as a Solver}
Solve phase of AMG is compute intensive and it involves computations of multigrid cycle over hierarchy of matrices. 
Further, data transfer between CPU's DRAM and GPU's device memory is often the most time consuming task and nullifies most of the speedups obtained from GPU. The latency due CPU-GPU  data transfer must be hidden to get a maximum gain from GPU computations. Further, GPU device memory needs to be managed efficiently, especially when it is a shared resource with several CPU cores. Taking these points into consideration, two hybrid algorithms are designed:
\begin{itemize}
 \item GPU-Compute Intensive (CPU/GPU-CI) algorithm
 \item GPU-Memory Intensive (CPU/GPU-MI) algorithm
\end{itemize}

		\begin{figure} 
			\centering
			\tikzset{every picture/.style={line width=0.75pt}} 
			
			\begin{tikzpicture}[x=0.75pt,y=0.75pt,yscale=-1,xscale=1,scale=1.0]
			
			\draw  [fill={rgb, 255:red, 0; green, 127; blue, 255 }  ,fill opacity=0.15 ] (149.5,78) -- (306.75,78) -- (306.75,120.5) -- (149.5,120.5) -- cycle ;
			\draw  [fill={rgb, 255:red, 138; green, 3; blue, 255 }  ,fill opacity=0.12 ] (148.5,38) -- (305.75,38) -- (305.75,60) -- (148.5,60) -- cycle ;
			\draw  [fill={rgb, 255:red, 255; green, 47; blue, 3 }  ,fill opacity=0.11 ] (305.75,38) -- (468.5,38) -- (468.5,60) -- (305.75,60) -- cycle ;
			\draw  [fill={rgb, 255:red, 53; green, 237; blue, 11 }  ,fill opacity=0.25 ] (306.75,78) -- (470.5,78) -- (470.5,120.5) -- (306.75,120.5) -- cycle ;
			\draw  [fill={rgb, 255:red, 0; green, 127; blue, 255 }  ,fill opacity=0.15 ] (306.75,120.5) -- (470.5,120.5) -- (470.5,159) -- (306.75,159) -- cycle ;
			\draw  [fill={rgb, 255:red, 53; green, 237; blue, 11 }  ,fill opacity=0.25 ] (149.5,120.5) -- (306.75,120.5) -- (306.75,159) -- (149.5,159) -- cycle ;
			\draw  [fill={rgb, 255:red, 243; green, 255; blue, 3 }  ,fill opacity=0.11 ] (150.5,168.5) -- (470.5,168.5) -- (470.5,227.5) -- (150.5,227.5) -- cycle ;
			\draw  [fill={rgb, 255:red, 0; green, 127; blue, 255 }  ,fill opacity=0.15 ] (151.5,250) -- (305.25,250) -- (305.25,289) -- (151.5,289) -- cycle ;
			\draw  [fill={rgb, 255:red, 53; green, 237; blue, 11 }  ,fill opacity=0.25 ] (305.25,250) -- (469.5,250) -- (469.5,289) -- (305.25,289) -- cycle ;
			\draw  [fill={rgb, 255:red, 53; green, 237; blue, 11 }  ,fill opacity=0.25 ] (151.5,289) -- (305.25,289) -- (305.25,326) -- (151.5,326) -- cycle ;
			\draw  [fill={rgb, 255:red, 0; green, 127; blue, 255 }  ,fill opacity=0.15 ] (305.5,289) -- (469.5,289) -- (469.5,326) -- (305.5,326) -- cycle ;
			\draw  [fill={rgb, 255:red, 0; green, 127; blue, 255 }  ,fill opacity=0.15 ] (149.5,334) -- (470.5,334) -- (470.5,361) -- (149.5,361) -- cycle ;
			\draw  [fill={rgb, 255:red, 0; green, 127; blue, 255 }  ,fill opacity=0.15 ] (147.5,15) -- (260.5,15) -- (260.5,35) -- (147.5,35) -- cycle ;
			\draw  [fill={rgb, 255:red, 53; green, 237; blue, 11 }  ,fill opacity=0.25 ] (265.75,15) -- (358.5,15) -- (358.5,35) -- (265.75,35) -- cycle ;
			\draw  [fill={rgb, 255:red, 243; green, 255; blue, 3 }  ,fill opacity=0.11 ] (364.5,15) -- (468.5,15) -- (468.5,35) -- (364.5,35) -- cycle ;
			\draw (230.25,49) node  [font=\normalsize] [align=left] {{\fontfamily{ptm}\selectfont\textcolor[rgb]{0,0,0}{ CUDA Stream 1}}};
			\draw (385.38,50) node  [font=\normalsize] [align=left] {{\fontfamily{ptm}\selectfont\textcolor[rgb]{0,0,0}{ CUDA Stream 2}}};
			\draw (302,72) node  [font=\footnotesize] [align=left] {{\fontfamily{ptm}\selectfont\textcolor[rgb]{0,0,0}{ for i = 0 to nlevels-2}}};
			\draw (236.13,98.25) node  [font=\footnotesize] [align=left] {{\fontfamily{ptm}\selectfont \textcolor[rgb]{0,0,0}{Perform Smoothing}}\\{\fontfamily{ptm}\selectfont \textcolor[rgb]{0,0,0}{of A[i],u[i],f[i] on GPU}\textcolor[rgb]{0,0,0}{}}};
			\draw (388.63,99.25) node  [font=\footnotesize] [align=left] {{\fontfamily{ptm}\selectfont \textcolor[rgb]{0,0,0}{Transfer A[i+1]}}\\{\fontfamily{ptm}\selectfont\textcolor[rgb]{0,0,0}{ ,P[i] to GPU}}};
			\draw (218.75,140) node  [font=\footnotesize] [align=left] {{\fontfamily{ptm}\selectfont\textcolor[rgb]{0,0,0}{ Transfer u[i] to CPU}}\\{\fontfamily{ptm}\selectfont \textcolor[rgb]{0,0,0}{ Initialize u[i+1]}}};
			\draw (388.63,139.75) node  [font=\footnotesize] [align=left] {{\fontfamily{ptm}\selectfont \textcolor[rgb]{0,0,0}{ Compute f[i+1] to GPU}}\\{\fontfamily{ptm}\selectfont \textcolor[rgb]{0,0,0}{ Transfer it to CPU}}};
			\draw (317.25,197) node  [font=\footnotesize] [align=left] {{\fontfamily{ptm}\selectfont\textcolor[rgb]{0,0,0}{ Coarse Level Direct Solve A[nlevels-1]}}\\{\fontfamily{ptm}\selectfont \textcolor[rgb]{0,0,0}{,u[nlevels-1], f[nlevels-1] on CPU}}\\{\fontfamily{ptm}\selectfont \textcolor[rgb]{0,0,0}{Transfer u[nlevels-1] to GPU and Prolongate}}};
			\draw (305,243) node  [font=\footnotesize] [align=left] {{\fontfamily{ptm}\selectfont \textcolor[rgb]{0,0,0}{ for i = levels-2 to 1}}};
			\draw (228.38,269.5) node  [font=\footnotesize] [align=left] {{\fontfamily{ptm}\selectfont\textcolor[rgb]{0,0,0}{Perform Smoothing}}\\{\fontfamily{ptm}\selectfont\textcolor[rgb]{0,0,0}{ of A[i],u[i],f[i] on GPU}}};
			\draw (377.25,269.75) node  [font=\footnotesize] [align=left] {{\fontfamily{ptm}\selectfont\textcolor[rgb]{0,0,0}{ Transfer P[i-1],u[i-1]} }\\{\fontfamily{ptm}\selectfont\textcolor[rgb]{0,0,0}{ to GPU}}};
			\draw (387.5,307) node  [font=\footnotesize] [align=left] {{\fontfamily{ptm}\selectfont \textcolor[rgb]{0,0,0}{ Prolongate u[i] }}};
			\draw (228.38,307.5) node  [font=\footnotesize] [align=left] {{\fontfamily{ptm}\selectfont\textcolor[rgb]{0,0,0}{ Transfer A[i-1] \& f[i-1] }}\\{\fontfamily{ptm}\selectfont\textcolor[rgb]{0,0,0}{ to GPU \ }}};
			\draw (317.88,348.25) node  [font=\footnotesize] [align=left] {{\fontfamily{ptm}\selectfont \textcolor[rgb]{0,0,0}{Perform Smoothing of A[0],u[0],f[0] on GPU}\textcolor[rgb]{0,0,0}{}}};
			\draw (203.63,25) node  [font=\footnotesize] [align=left] {{\small {\fontfamily{ptm}\selectfont\textcolor[rgb]{0,0,0}{ Compute on GPU}}}};
			\draw (312.13,25) node  [font=\footnotesize] [align=left] {{\small {\fontfamily{ptm}\selectfont \textcolor[rgb]{0,0,0}{ Data Transfer}}}};
			\draw (416.5,25) node  [font=\footnotesize] [align=left] {{\small {\fontfamily{ptm}\selectfont\textcolor[rgb]{0,0,0}{ Compute on CPU}}}};
			\end{tikzpicture}
			\caption{Data transfer scheme for Hybrid CPU/GPU-CI implementation}
			\label{gpu-transfer}
		\end{figure}
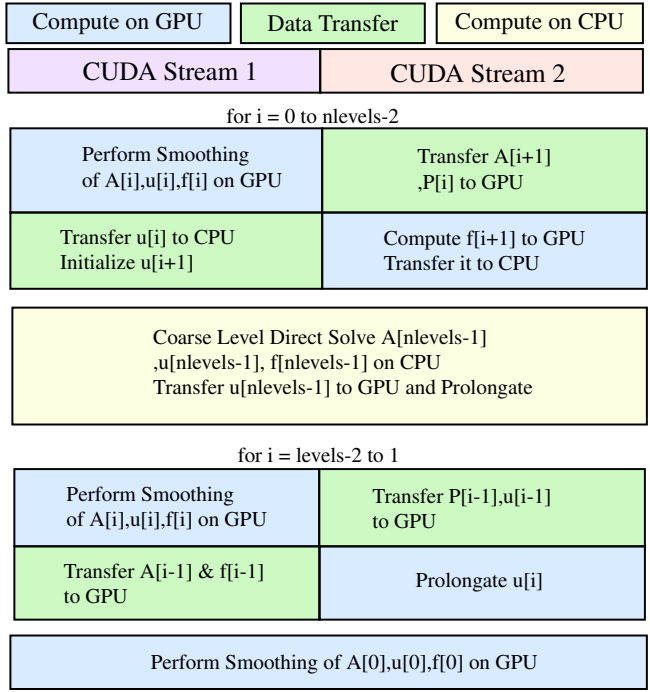 
		
\subsubsection{Hybrid CPU/GPU-CI Algorithm}		
 CPU/GPU-CI algorithm significantly reduces GPU memory requirements and hides the CPU-GPU data transfer latency by overlapping data transfer with GPU computations. Further, it exploits the fact that   at any instant AMG computations  need data only from one hierarchy  level. Therefore, GPU memory requirement can be reduced by keeping only the matrices involved in computations  and in data transfer. All other matrices can be stored in CPU and transferred only when needed by GPU.
		
Initially the system matrix and its RHS, the system at hierarchy  $Level:i=0$  is transferred to GPU and then the pre-smoothing iteration is initiated in CUDA Stream~1. Simultaneously, the transfer of the prolongation matrix at $Level:i$ and the  coarse matrix at $Level:i+1$ to GPU are initiated in CUDA Stream~2.
Here, the smoothing iteration in CUDA Stream~1   overlaps with the matrix transfer  in CUDA Stream~2 and it hide the data transfer latency.
Then, the smoothened solution of $Level:i$ is transferred to CPU's DRAM. During this transfer, the residual is restricted to get RHS of $Level:i+1$ and transferred back to CPU by the other CUDA stream. On reaching the  second last level,  the residual is restricted to form RHS of last level and it is transferred back to CPU. A direct solver is used to solve the coarsest level system with OpenMP parallelization. The solution of coarsest level is then transferred back to GPU for prolongation. 
The process is repeated while propagating from coarsest to finest level. 
Fig.~\ref{gpu-transfer} shows the   overlapped transfer and computations performed on CPU and GPU. Although data movement between CPU and GPU is augmented by this approach, most of its latencies are hidden by utilizing GPU on compute intensive smoothing operations at the same time.   Moreover, this algorithm requires significantly less device memory compared to GPU-only implementations, see the numerical experiment section.     \\    
		
\noindent\textbf{Remark:}		
The proposed data transfer scheme can be applied to any multigrid cycle in a single node or in a distributed system. 
		
\subsubsection{Hybrid CPU/GPU-MI}
This algorithm is designed to further improve the performance by utilizing more device memory in comparison to Hybrid CPU/GPU-CI. In this approach,  the system matrices of all but coarsest level,  and prolongation matrices are stored on GPU. Coarsest level matrix is LU factorized during the setup phase and the LU factorization requires a large amount of memory to store the factors $L$ and $U$. Therefore,  the coarsest level system is solved   on CPU using OpenMP parallelization as in CPU/GPU-CI algorithm.

\subsection{AMG as a Preconditioner}
The number of iterations in Krylov subspace methods is reduced drastically when AMG is used as  a preconditioner. 
Since AMG does not require access to the physical grids to build a hierarchy of matrices, it can be used as a black box preconditioner in any of the Krylov subspace methods.
Conjugate Gradient (CG) method and Biconjugate Gradient Stablilized (BiCG) methods are considered to evaluate the performance of AMG as a preconditioner.

\begin{algorithm}
\caption{Preconditioned Conjugate Gradient Algorithm}
\label{PCG}
\begin{algorithmic}[1]
\STATE{\textbf{INPUT:} $A$: $n$ $\times$ $n$ sparse matrix; $b$ RHS vector, $tol$ tolerance, Preconditioner $M$}

\STATE{\textbf{OUTPUT:} $x$ Solution Vector}

\STATE{	Compute $r_{0}$ = $b$ - $Ax_{0}$ , $z_{0}$ = $M^{-1}$$r_{0}$\null\hfill \textnormal{{Perform one AMG cycle on $Az_{0}=r_{0}$}}\\}
\STATE{$p_{0}$ = $z_{0}$\\} 
\FOR{j = 0,1,.... till $\mid\mid r_{j} \mid\mid > tol$ }
    \STATE{$\alpha_{j}$ = $(r_{j},z_{j})/(Ap_{j},p_{j})$\\}
    \STATE{$x_{j+1}$ = $x_{j}$ + $\alpha_{j}p_{j}$ \\}
    \STATE{$r_{j+1}$ = $r_{j}$ + $\alpha_{j}Ap_{j}$ \\}
    \STATE{$z_{j+1}$ = $M^{-1}$$r_{j+1}$\null\hfill \textnormal{{Perform one AMG cycle on $Az_{j+1}=r_{j+1}$}}\\}
    \STATE{$\beta_{j}$ = $(r_{j+1},z_{j+1})/(r_{j},z_{j})$\\}
    \STATE{$p_{j+1}$ = $z_{j+1}$ + $\beta_{j}p_{j}$}
\ENDFOR
\RETURN $x$ vector
\end{algorithmic}
\end{algorithm} 

Algorithm \ref{PCG} highlights the steps involved in AMG Preconditioned Conjugate Gradient (AMG-PCG) solver~\cite{saad2003iterative}. Computations involved in  each iteration of PCG-AMG algorithm are divided into two groups: $(i)$ AMG preconditioning Step 7 of algorithm \ref{PCG} and $(ii)$   the remaining steps in PCG, i.e. Steps 4-6 and 8,~9 referred to as CG steps. Four variants of AMG PCG are implemented which are as follows:
\begin{enumerate}
	\item \textbf{AMG-PCG 1:} All   computations are performed on CPU with OpenMP multi-threaded setting. This variant is considered as a baseline for comparisons.
	\item \textbf{AMG-PCG 2:} AMG preconditioning   is performed using Hybrid CPU/GPU-CI algorithm and CG steps are executed on CPU with OpenMP multi-threaded setting.
	\item \textbf{AMG-PCG 3:} Both AMG preconditioning    using Hybrid CPU/GPU-CI  algorithm and CG steps are executed  on GPU.
	\item \textbf{AMG-PCG 4:} AMG preconditioning   using  Hybrid CPU/GPU-MI algorithm  and CG steps are  executed on GPU.
\end{enumerate}

\begin{algorithm}
\caption{Preconditioned Flexible BiCG Algorithm}
\label{PBiCG}
\begin{algorithmic}[1]
\STATE{\textbf{INPUT:} $A$: $n$ $\times$ $n$ sparse matrix; $b$ RHS vector, $tol$ tolerance, Preconditioner $M$, $\overline{r_{0}}$ arbitrary}

\STATE{\textbf{OUTPUT:} $x$ Solution Vector}

\STATE{	Compute $r_{0}$ = $b$ - $Ax_{0}$ , $\overline{r_{0}}$ arbitrary\\}
\FOR{j = 0,1,.... till $\mid\mid r_{j} \mid\mid > tol$ }
    \STATE{$\widetilde{p_{j}}$ = $M^{-1}p_{j}$\null\hfill \textnormal{{Perform one AMG cycle on $A\widetilde{p_{j}}=p_{j}$}}\\}
    \STATE{$\alpha_{j}$ = $(r_{j},\overline{r_{0}})/(A\widetilde{p_{j}},\overline{r_{0}})$\\}
    \STATE{$s_{j}$ = $r_{j}$ - $\alpha_{j}A\widetilde{p_{j}}$ \\}
    \STATE{$\widetilde{s}_{j}$ = $M^{-1}s_{j}$ \null\hfill \textnormal{{Perform one AMG cycle on $A\widetilde{s}_{j}=s_{j}$}}\\}
    \STATE{$\omega_{j}$ = ($A\widetilde{s}_{j},s_{j})/(A\widetilde{s}_{j},A\widetilde{s}_{j})$\\}
    \STATE{$x_{j+1}$ = $x_{j}$ + $\alpha_{j}\widetilde{p}_{j}$ + $\omega_{j}\widetilde{s}_{j}$\\}
    \STATE{$r_{j+1}$ = $s_{j}$ - $\omega_{j}A\widetilde{s}_{j}$ \\}
    \STATE{$\beta_{j}$ = $(r_{j+1},\overline{r_{0}})/(r_{j},\overline{r_{0}})$ $\cdot$   ($\alpha_{j}/\omega_{j}$)\\}
    \STATE{$p_{j+1}$ = $r_{j+1}$ + $\beta_{j}(p_{j}-\omega_{j}A p_{j})$\\}
\ENDFOR
\RETURN $x$ vector
\end{algorithmic}
\end{algorithm}
BiCG solver does not require matrix to be symmetric and hence can handle larger class of problems. Chen $et.$ $al.$ \cite{chen2016analysis} highlight the application of preconditioned BiCG algorithm for solving linear systems. Each iteration of algorithm \ref{PBiCG} with AMG as preconditioner involves computation of two AMG cycles, one at Step~5 and other at Step~8 of the algorithm. Similar to AMG-PCG, four variants of AMG-preconditioned BiCG (AMG-PBiCG) algorithms, namely \textbf{AMG-PBiCG 1, AMG-PBiCG~2, AMG-PBiCG~3, AMG-PBiCG 4} are implemented. 

\section{Numerical Experiments}
\label{sec_experiments}
We first analysis the performance of different coarsening algorithms and then study  the efficiency of the proposed hybrid AMG as a solver and as a preconditioner to Krylov solvers. For this analysis, symmetric and unsymmetric matrices with varying sparsity pattern are used. After that the performance of the hybrid implementations is compared with AMGX, the GPU-only implementation, which is specifically designed to exploit GPU architectures. More general matrices given in Sparse Suite collection~\cite{davis2011university} are also used in the comparative study.

All experiments are performed on a workstation equipped with Intel Xeon Gold 6150 with base clock-speed of 2.7 GHz with 3.7 GHz Turbo boost, 18 cores with hyper-threading enabled, 192GB RAM (16GB$\times$12) and NVIDIA GV100 GPU with 32 GB device memory, 5120 CUDA threads. Intel C++ 19.1 compiler with Intel MKL Sparse BLAS library \cite{wang2014intel} and CUDA 10.2 compiler with CUSPARSE library \cite{naumov2010cusparse} are used. Further, the coarsest level system in AMG is solved with a direct solver PARDISO~\cite{schenk2004solving}. CPU implementation with OpenMP parallelism forms a baseline implementation for evaluating performance of hybrid approaches, where 16 OpenMP threads are considered in computations. Further, the stopping criteria in all experiments  
is prescribed as $\mid\mid b-Ax \mid\mid_{2}$ $<$ $10^{-8}$.

\subsection{Evaluation of coarsening algorithms}
Matrices to evaluate different coarsening algorithms are obtained from the scalar convection, diffusion, reaction equation 
\[
-\Delta u + \mathbf{b}\cdot\nabla u + cu  = f \quad \text{in} \quad (0,1)^2
\]
with an inhomogeneous Dirichlet boundary condition. Matrices with different sparsity patterns are obtained by discretizing the scalar equation with different orders of finite element (FE) on a triangulated domain from the in-house finite element package ParMooN~\cite{PAR01,wilbrandt2017parmoon}. Moreover,
symmetric and unsymmetric matrices are obtained with $\mathbf{b}=\mathbf{0}$, $c=0$ and $\mathbf{b}={(1,100)}^T$, $c=1$, respectively.   The obtained matrix types are given in Table~\ref{tab_1}.
\begin{table}
	\caption{Types of matrices obtained by discretizing the scalar equation with different orders of finite element  ($P_1-P_4$) and by uniformly refining the mesh (L1-L4).}
	\centering
    \begin{tabular}{ccrrc}
		\hline
		\textbf{FE order}    & \textbf{Matrix} &  \textbf{Size}  & \textbf{Non-zeros} & \textbf{Sparsity}   \\ 
		\hline
		                    &P1L1& 97,537   & 679,705   & 7.14E-05 \\ 
	  $P_1$      &P1L2& 391,681  & 2,735,641  & 1.78E-05 \\
							&P1L3& 1,569,793 & 10,976,281 & 4.45E-06 \\
							&P1L4& 6,285,313 & 43,972,633 & 1.11E-06 \\		
		\hline 
					 		&P2L1& 97537   & 1112145  & 1.17E-04 \\ 
		 $P_2$		& P2L2& 391681  & 4485201  & 2.92E-05 \\
							& P2L3& 1569793 &	18014289 & 7.31E-06 \\
							& P2L4& 6285313 & 72204369 & 1.83E-06 \\ 
		\hline
							& P3L1 & 54721   & 916585   & 3.06E-04 \\ 
		 $P_3$ 	    & P3L2& 220033  & 3713065  & 7.67E-05 \\ 
							& P3L3& 882433  & 14946217 & 1.92E-05 \\ 
							& P3L4& 3534337 & 59973289 & 4.80E-06 \\
		\hline 
		 					& P4L1& 97537   & 2262817  & 2.38E-04 \\ 
		 $P_4$      & P4L2& 391681  & 9145633  & 5.96E-05 \\ 
							& P4L3& 1569793 & 36772129 & 1.49E-05 \\ 
							& P4L4& 6285313 & 147468577&	3.73E-06 \\
		\hline
	\end{tabular}
	\label{tab_1}
\end{table}
 
\begin{figure}
	\centering
	\begin{subfigure}{0.5\textwidth}
		\includegraphics[scale=0.4]{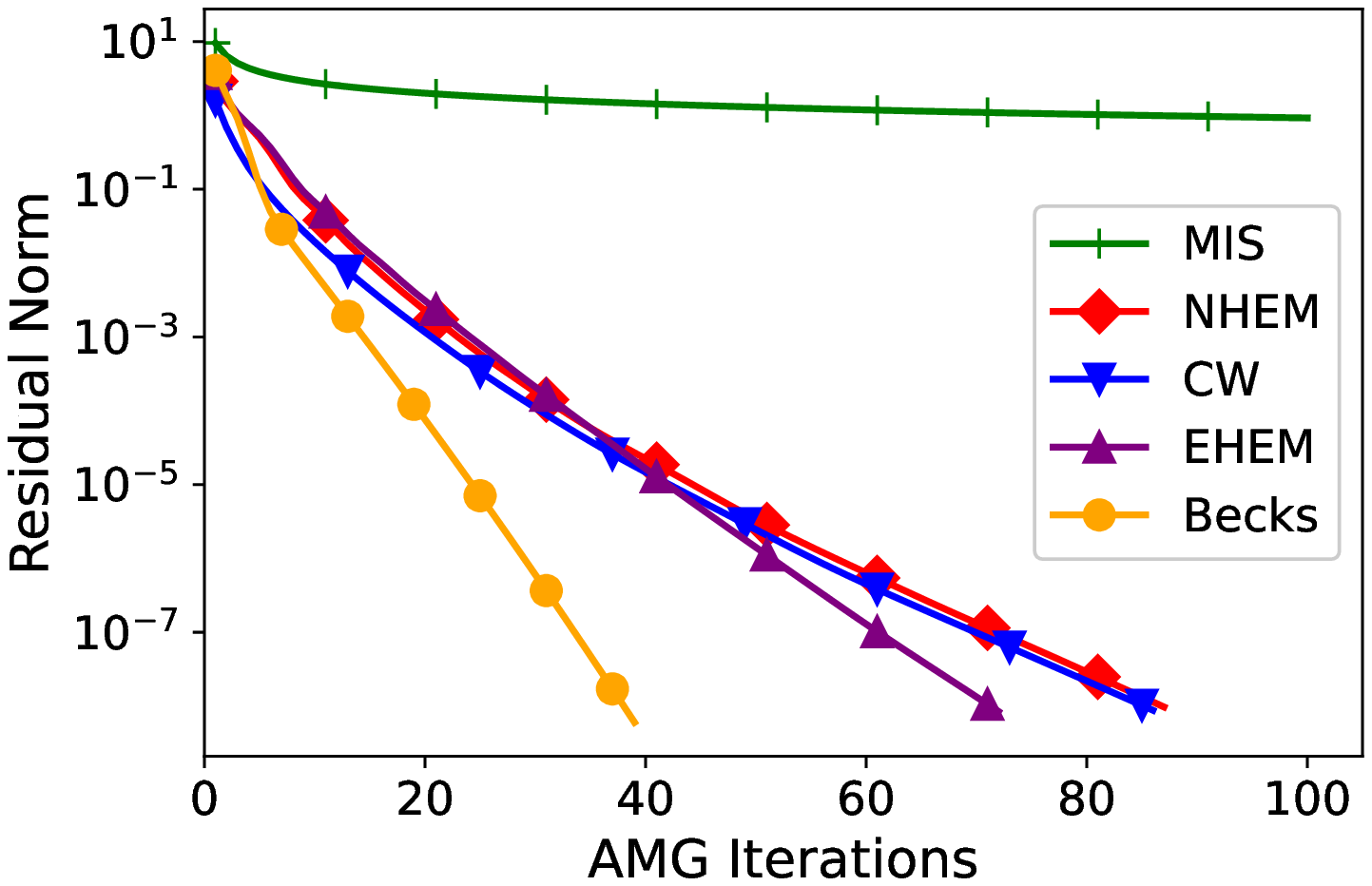}
		\subcaption{Comparison of convergence}
		\label{coarsen_1}
	\end{subfigure}%
	\begin{subfigure}{0.5\textwidth}
		\includegraphics[scale=0.4]{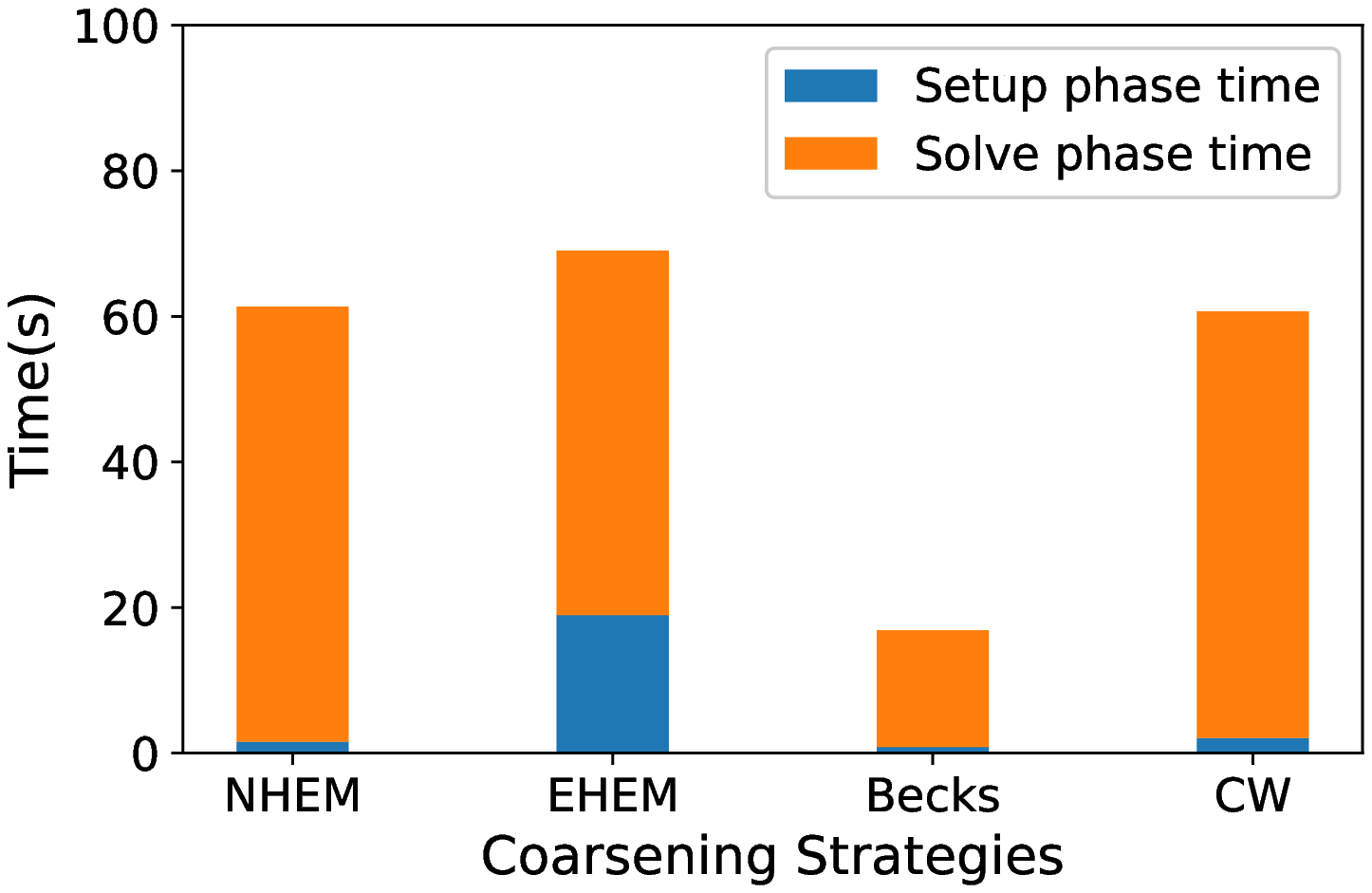}
		\subcaption{Time Comparison}
		\label{coarsen_2}
	\end{subfigure}
	\centering
	\vspace{-2mm}
	\caption{Comparison of coarsening strategies in AMG solver}
	\label{coarsen_12}
\end{figure}

Convergence properties of AMG is highly dependent on the type of coarsening algorithms used to construct hierarchy of matrices. In this analysis, we compare the following coarsening algorithms 
\begin{itemize}
    \item Beck's classical algorithm
    \item Node-base HEM with unsmoothed aggregation  (NHEM)
    \item Edge-base HEM with unsmoothed aggregation (EHEM)
    \item Node-base HEM with smoothed aggregation (compatible weighted matching) (CW)
    \item Maximal Independent Set (MIS) with unsmoothed aggregation~\cite{7397636}.
\end{itemize}
These coarsening strategies are compared in AMG solver and in AMG preconditioned Krylov subspace solvers.
An unsymmetric matrix of type P1L4 is used in AMG solver, whereas  a symmetric matrix of type P1L4 is used in AMG preconditioned Krylov subspace solvers. Moreover, these experiments are performed in a multi-threaded settings with 16 OpenMP threads, see Table~\ref{tab_3} for other parameters of AMG.
     
Fig.~\ref{coarsen_1} compares the convergence of the solution obtained from AMG solver with different coarsening algorithms. 
The convergence in MIS is very poor comparing to other coarsening algorithms. Since the coarsening ratio in MIS depends on the average degree of the graph, it varies on each hierarchical level.
Note that the average degree of the graph increases on every application of MIS coarsening, and consequently the coarsening ratio  also increases on coarse levels. Such higher coarsening ratio results in poor projections of vectors across the levels and results in poor convergence. 
NHEM, EHEM and CW coarsening algorithms have average coarsening ratio of two and hence they show similar convergence characteristics. Since the aggregation procedure, except the values of the prolongation matrix $P$, is same in NHEM and CW, the computing time is also similar in both approaches. The setup time in 
EHEM coarsening algorithm is significant compared to NHEM due to sorting operation of edge list, see fig~\ref{coarsen_12}. Nevertheless, EHEM coarsening is invariant to the numbering of DOFs and hence provides a consistent hierarchy of matrices even a matrix reordering is performed. 
\begin{table}
    \centering
	\caption{AMG Parameters}\label{tab_3}
	\begin{tabular}{cccc}
		\hline
		\textbf{AMG} & \textbf{Coarsest Level} & \textbf{Presmoothing}& \textbf{Postsmoothing} \\
		\textbf{Levels} & \textbf{Matrix Size} &  \textbf{iterations} & \textbf{iterations}\\
		\hline
		6 & 20,000-40,000 & 6 & 6\\ 
		\hline
	\end{tabular}
\end{table}
\begin{figure}
	\centering
	\begin{subfigure}{.5\textwidth}
		\centering
		\includegraphics[scale=0.4]{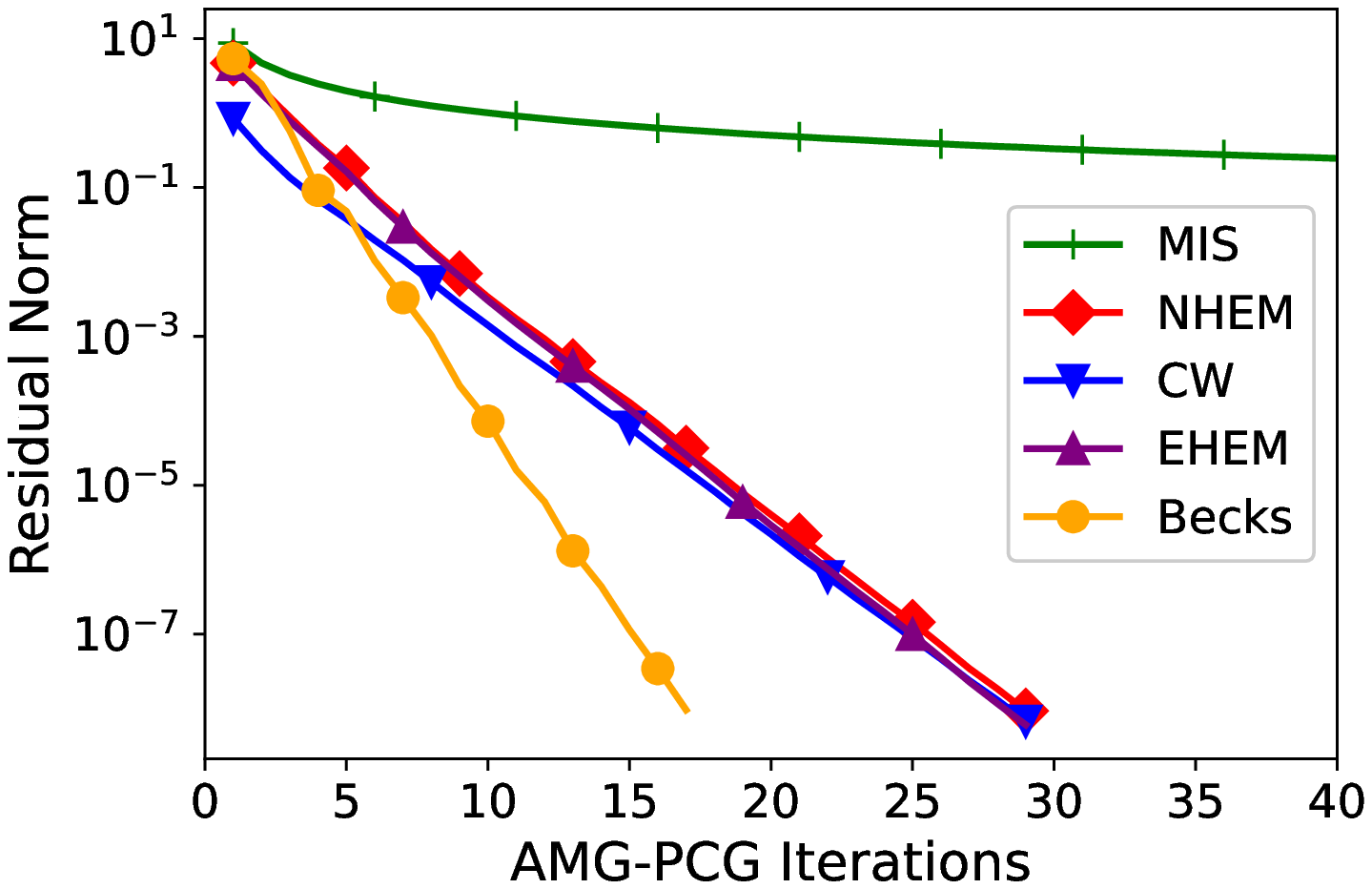}
		\vspace{-2mm}
		\caption{Comparison of convergence}
		\label{coarsen_3}
	\end{subfigure}%
	\begin{subfigure}{.5\textwidth}
		\includegraphics[scale=0.4]{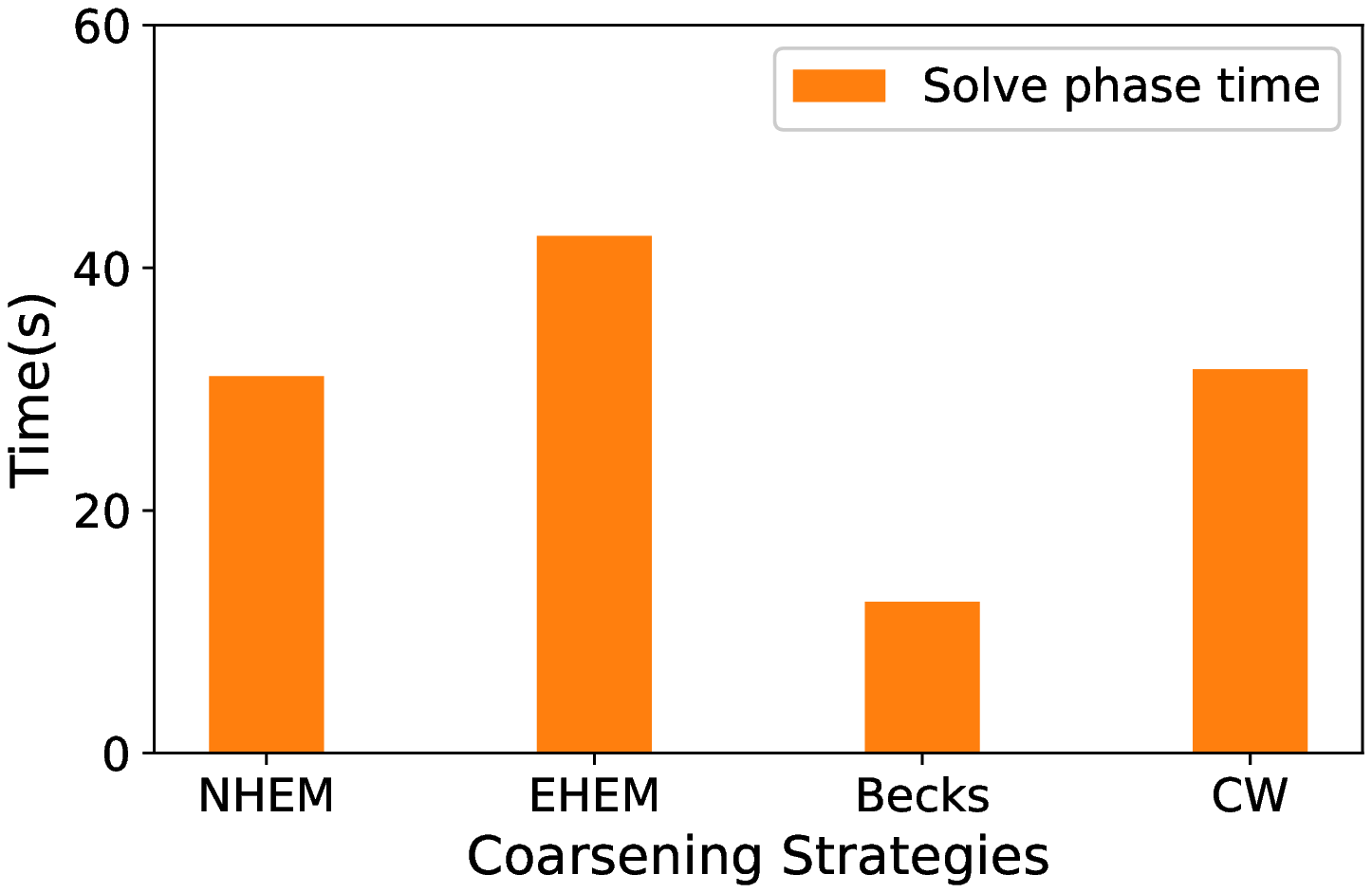}
		\vspace{-2mm}
		\caption{Time Comparison}
		\label{coarsen_4}
	\end{subfigure}
	\vspace{-2mm}
	\caption{Comparison of coarsening strategies in AMG-PCG}
	\label{coarsen_34}
\end{figure}

Next, we compare the performance of coarsening algorithms in AMG preconditioned Conjugate Gradient method (AMG-PCG) given in Algorithm \ref{PCG}.
Fig.~\ref{coarsen_3} and Fig.~\ref{coarsen_4} show the convergence characteristics and the time taken by the solve phase. The behaviour of different coarsening algorithms in AMG-PCG   is similar to the behaviours observed in AMG solver.

Overall, Beck's coarsening approach takes lowest computing time among all five coarsening strategies considered, see Fig.~\ref{coarsen_2}  and \ref{coarsen_4}. Its localized averaging approach does not consider matrix entries into consideration while coarsening and it results in lower setup time. Although anisotropic problems are in general challenging to handle  by classical coarsening algorithms~\cite{gandham2014gpu}, robust algorithms like Beck's can handle these class of problems efficiently. The efficiency of all variants of HEM coarsening algorithms is similar. Moreover, EHEM coarsening can be preferred when the influence of DOF numbering and/or matrix reordering need to be avoided.  
In all our further experiments, we use NHEM coarsening algorithm.


\subsection{Comparison of coarse level solver}
\begin{figure}[tbh]
	\centering
	\includegraphics[scale=0.4]{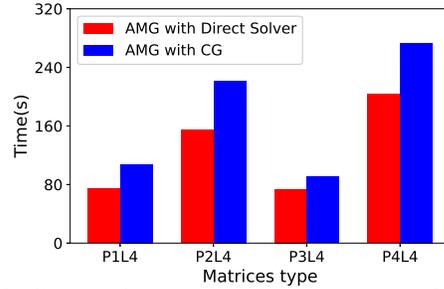}
	\vspace{-5mm}
	\caption{Comparison of coarse level solvers in AMG}
	\label{coarse_level_solver}
\end{figure}
Any direct or iterative solver with pre-defined number of smoothing iterations can be used as a coarsest level solver. 
In the present study, the performance of AMG Solver with CG and with direct solver PARDISO from MKL~\cite{wang2014intel} as a coarsest  solver is compared. Further, computations are performed for symmetric matrices of type P1L4, P2L4, P3L4 and P4L4. Fig.~\ref{coarse_level_solver} shows the total computing time taken in each computation. Setup phase of AMG with direct solver involves additional computation of LU factorization. Nevertheless, the direct solver takes less time in solve phase since it involves only forward and backward-substitution. Contrarily, CG takes more time in solve phase since the coarse system needs to be solved in every cycle of AMG. Hence, AMG with direct solver at coarsest  level  is recommended.         

\subsection{Complexity of AMG}
\begin{figure}
	\centering
	\begin{subfigure}{.5\textwidth}
		\centering
		\includegraphics[scale=0.4]{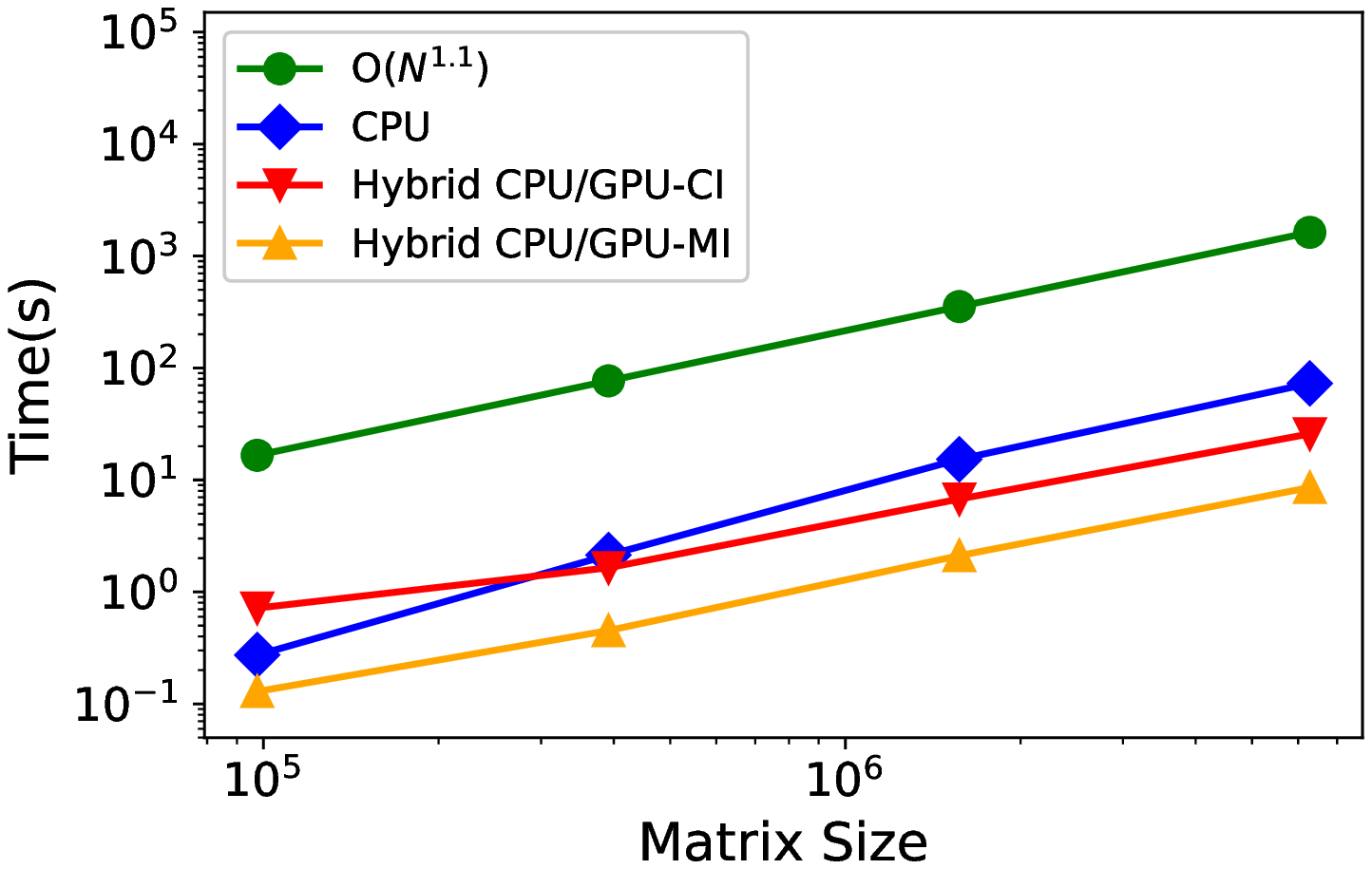}
		\vspace{-4mm}
		\caption{Complexity of AMG with P1 FE}
		\label{P1V1}
	\end{subfigure}%
	\begin{subfigure}{.5\textwidth}
		\centering
		\includegraphics[scale=0.4]{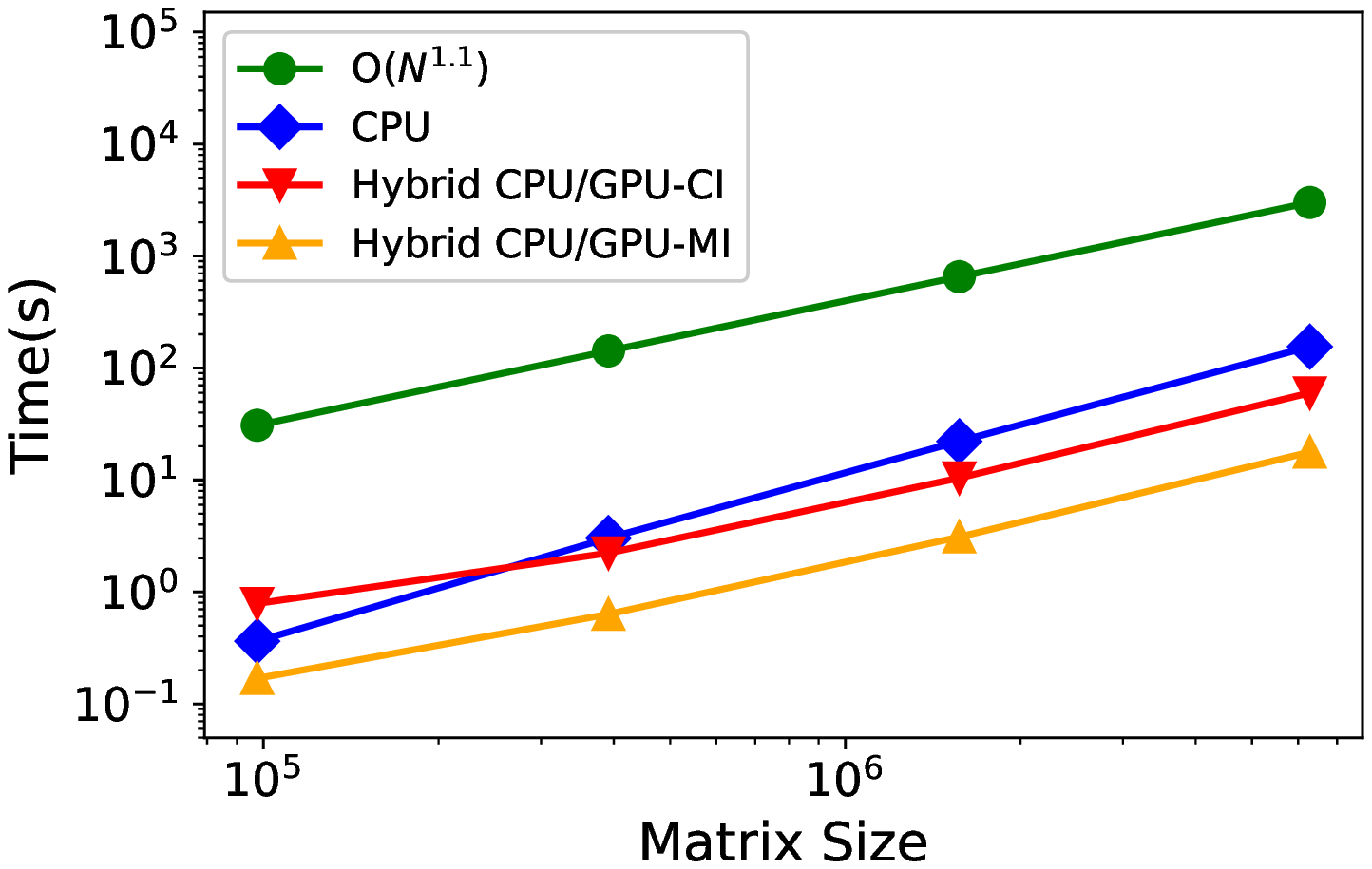}
		\vspace{-4mm}
		\caption{Complexity of AMG with P2 FE}
		\label{P1V2}
	\end{subfigure}
	\vspace{4mm}
	\begin{subfigure}{.5\textwidth}
		\centering
		\includegraphics[scale=0.4]{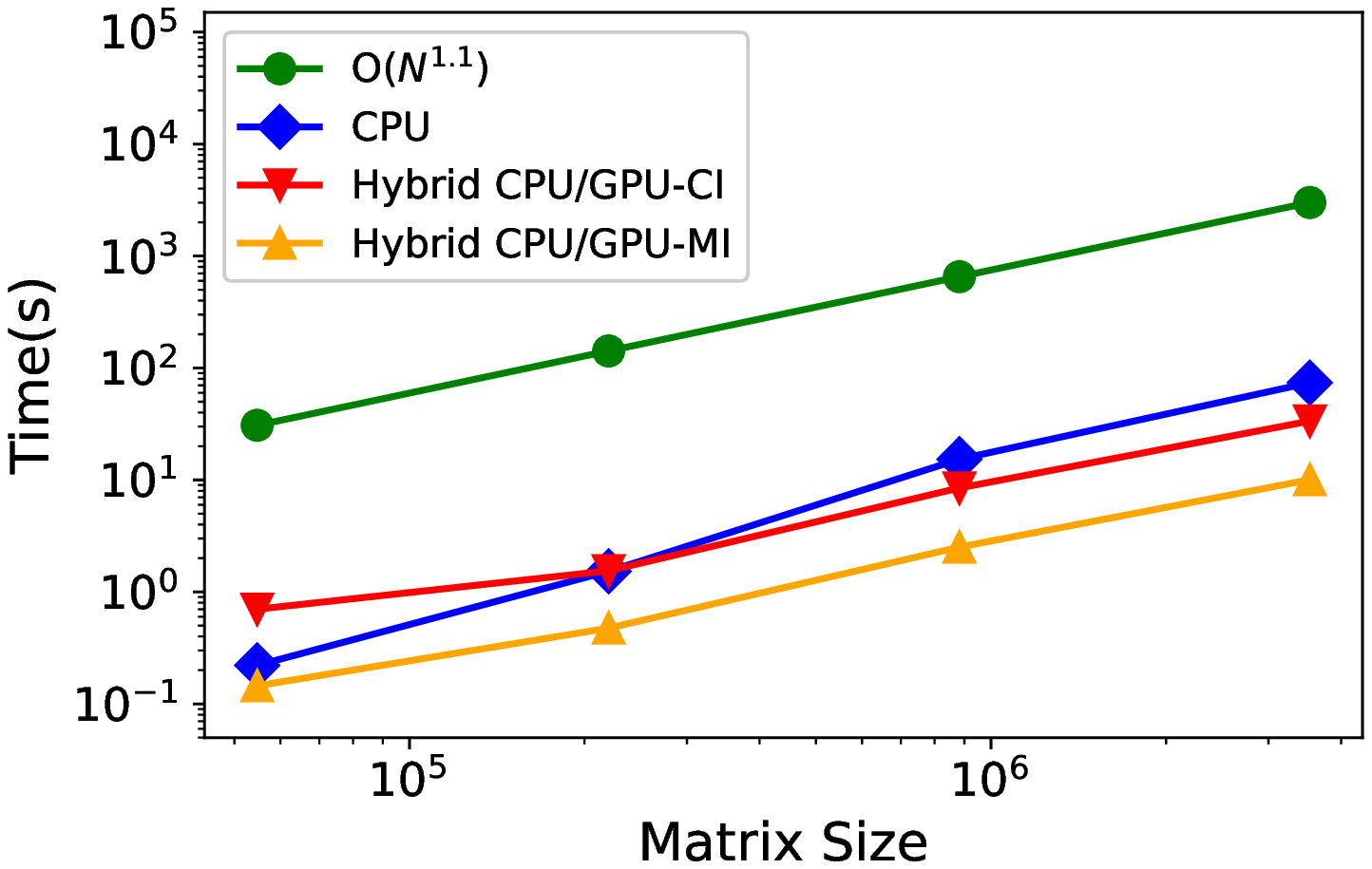}
		\vspace{-4mm}
		\caption{Complexity of AMG with P3 FE}
		\label{P1V3}
	\end{subfigure}%
	\begin{subfigure}{.5\textwidth}
		\centering
		\includegraphics[scale=0.4]{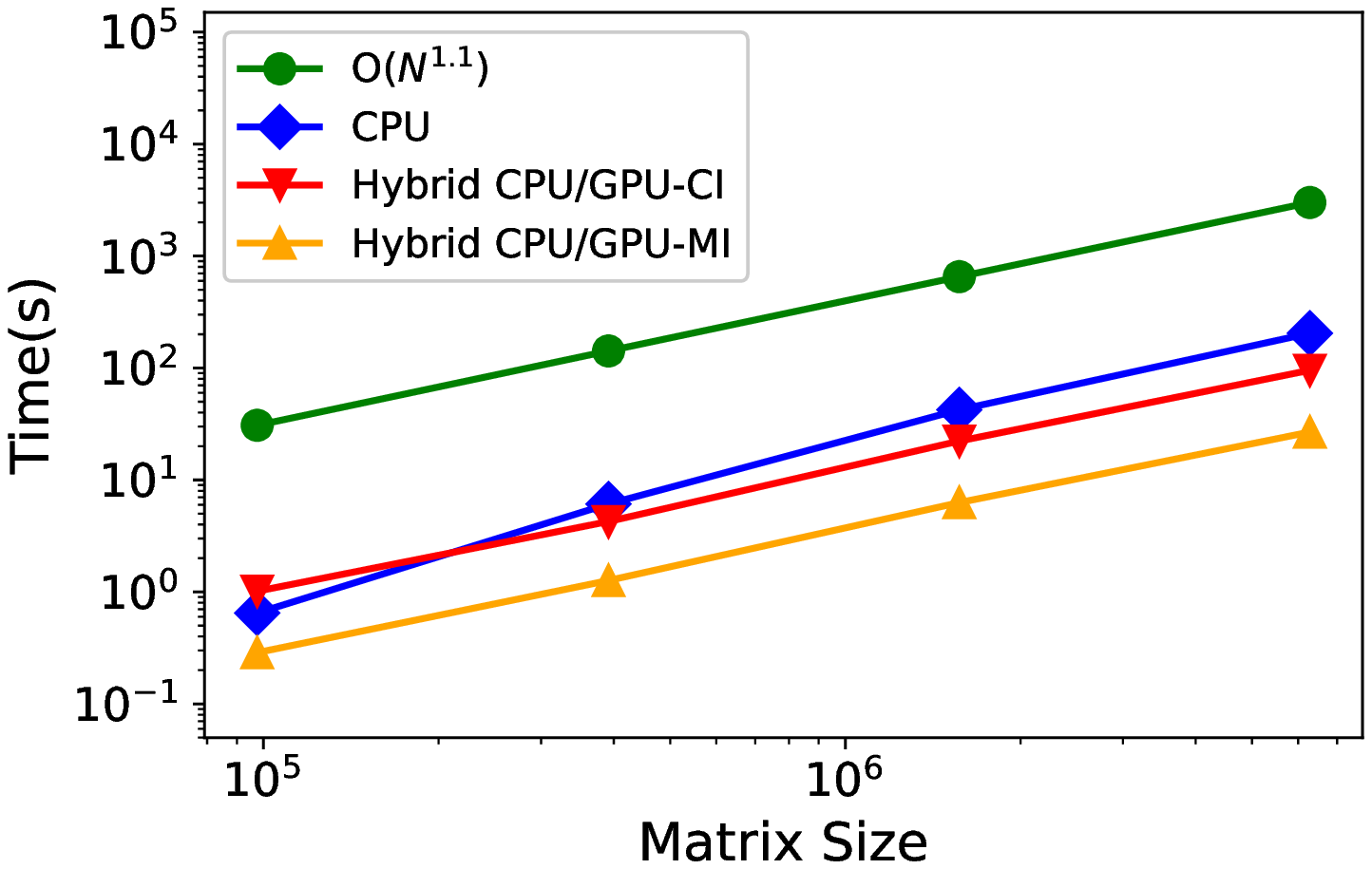}
		\vspace{-4mm}
		\caption{Complexity of AMG with P4 FE}
		\label{P1V4}
	\end{subfigure}
	\caption{Complexity of AMG solver for symmetric matrices}
	\label{P1V}
\end{figure}

Performance of sparse solvers is highly dependent on sparsity pattern of the matrix. In order to evaluate the complexity of AMG, matrices of different sizes but with same sparsity pattern and properties are needed. It is obtained by uniformly refining the mesh with same order of finite element.  Further, matrices with different sparsity patterns are considered to evaluate the complexity of AMG by varying the order of finite elements, see Table~\ref{tab_1}.
Hybrid CPU/GPU-CI and  Hybrid CPU/GPU-MI parallel implementations are compared with the baseline CPU implementation.

\begin{figure}
	\centering
	\begin{subfigure}{.5\textwidth}
		\centering
		\includegraphics[scale=0.4]{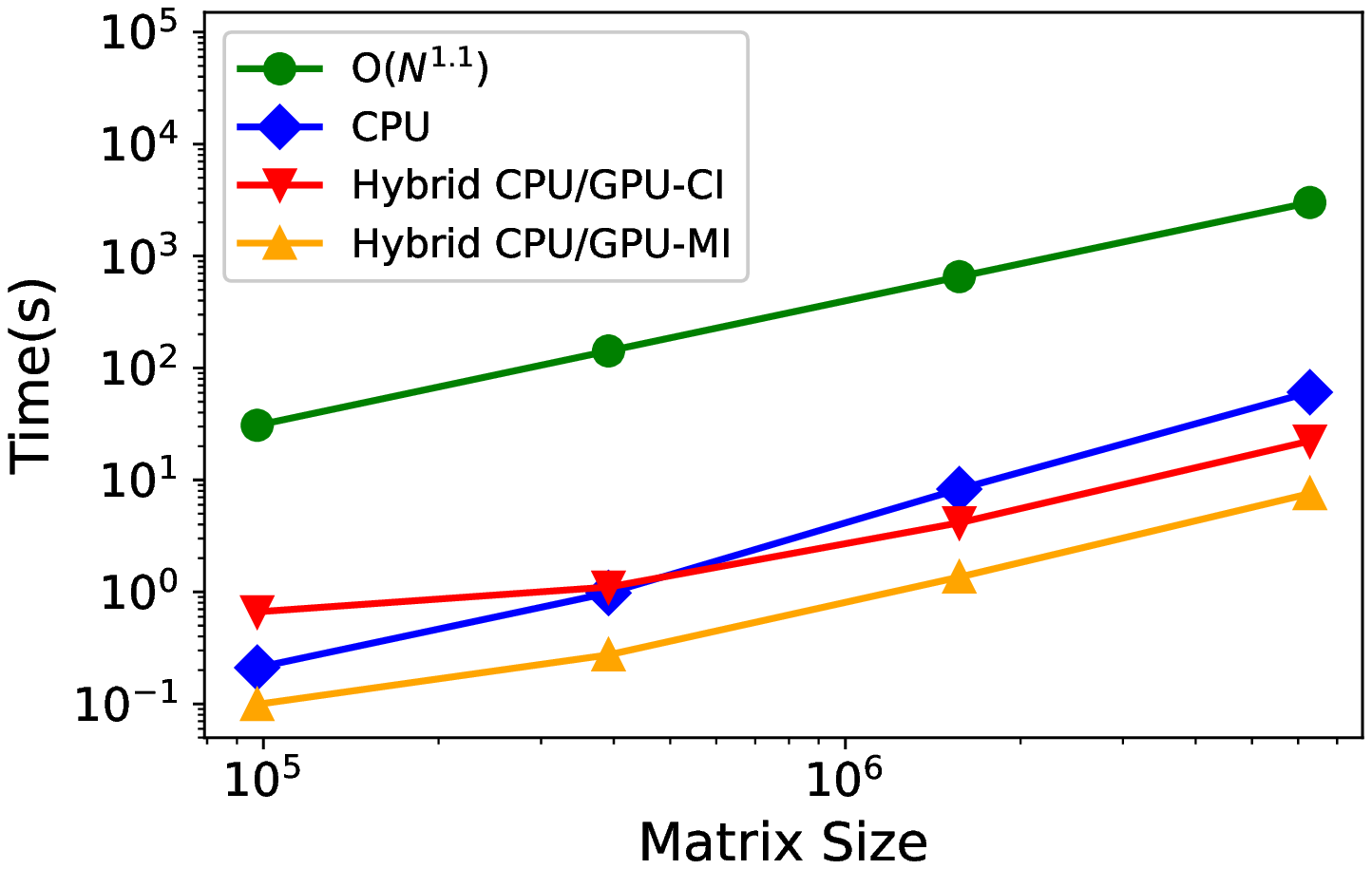}
		\vspace{-4mm}
		\caption{Complexity of AMG with P1 FE}
		\label{P2V1}
	\end{subfigure}%
	\begin{subfigure}{.5\textwidth}
		\centering
		\includegraphics[scale=0.4]{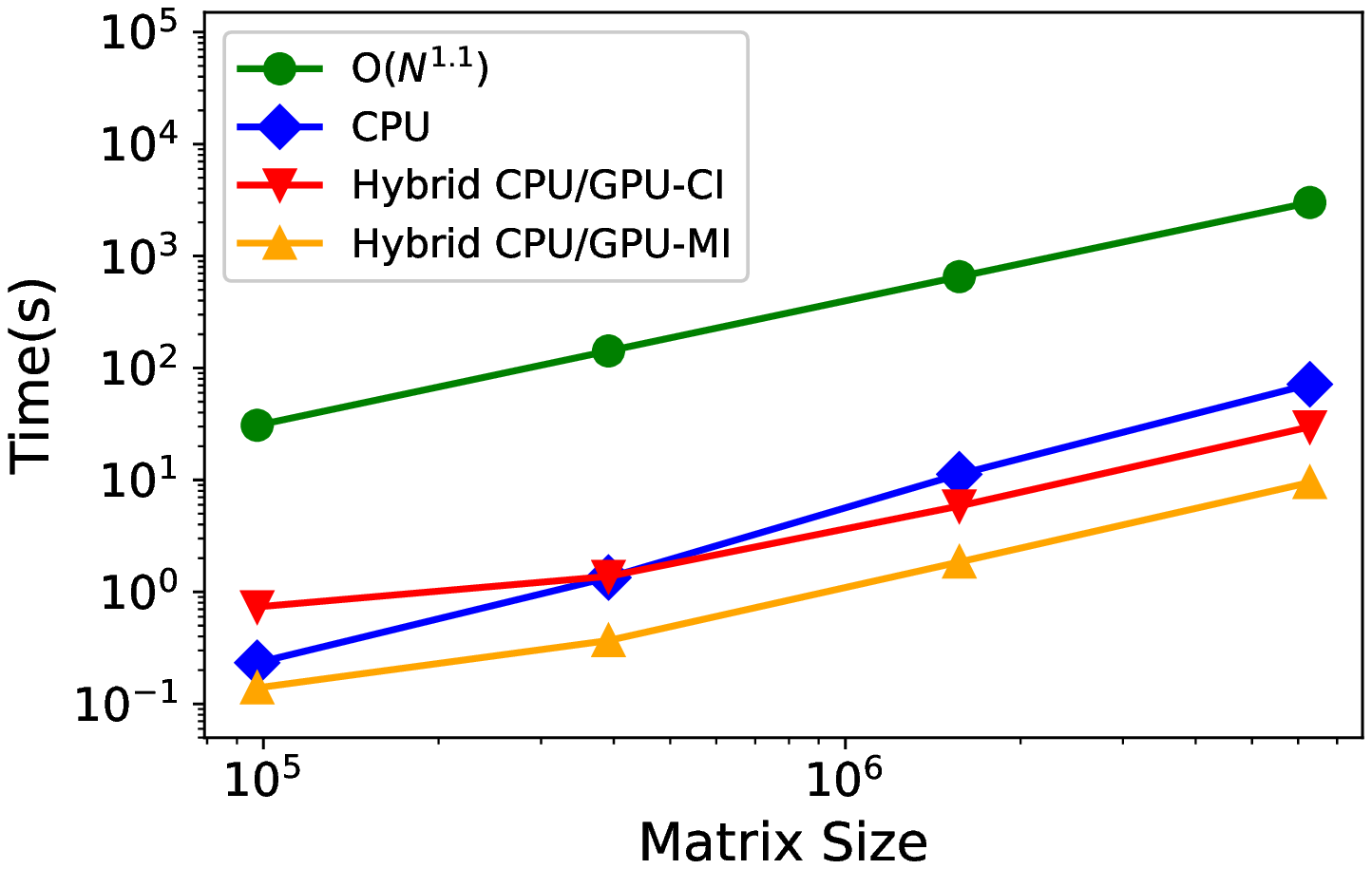}
		\vspace{-4mm}
		\caption{Complexity of AMG with P2 FE}
		\label{P2V2}
	\end{subfigure}
	\vspace{4mm}
	\begin{subfigure}{.5\textwidth}
		\centering
		\includegraphics[scale=0.4]{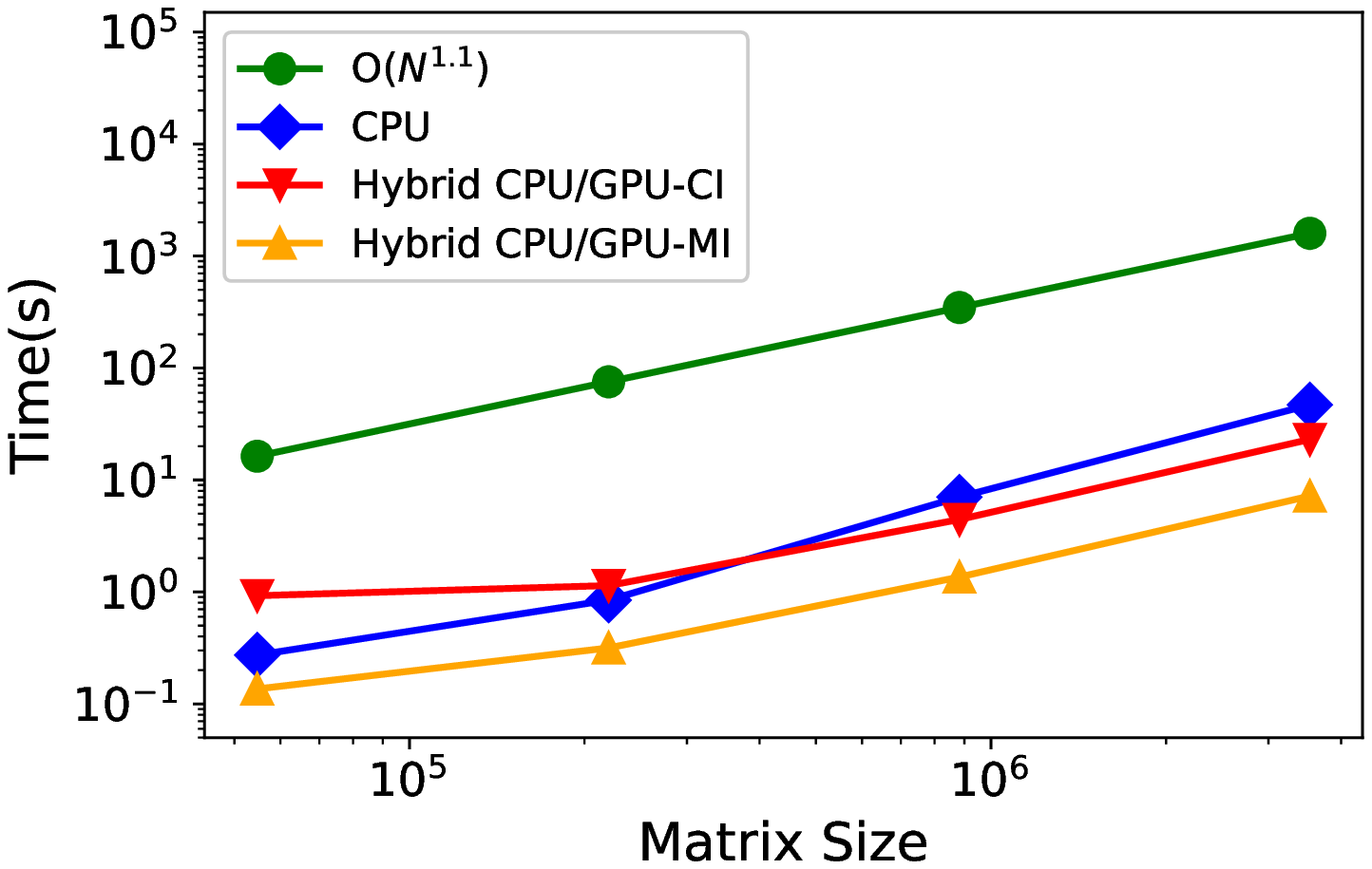}
		\vspace{-4mm}
		\caption{Complexity of AMG with P3 FE}
		\label{P2V3}
	\end{subfigure}%
	\begin{subfigure}{.5\textwidth}
		\centering
		\includegraphics[scale=0.4]{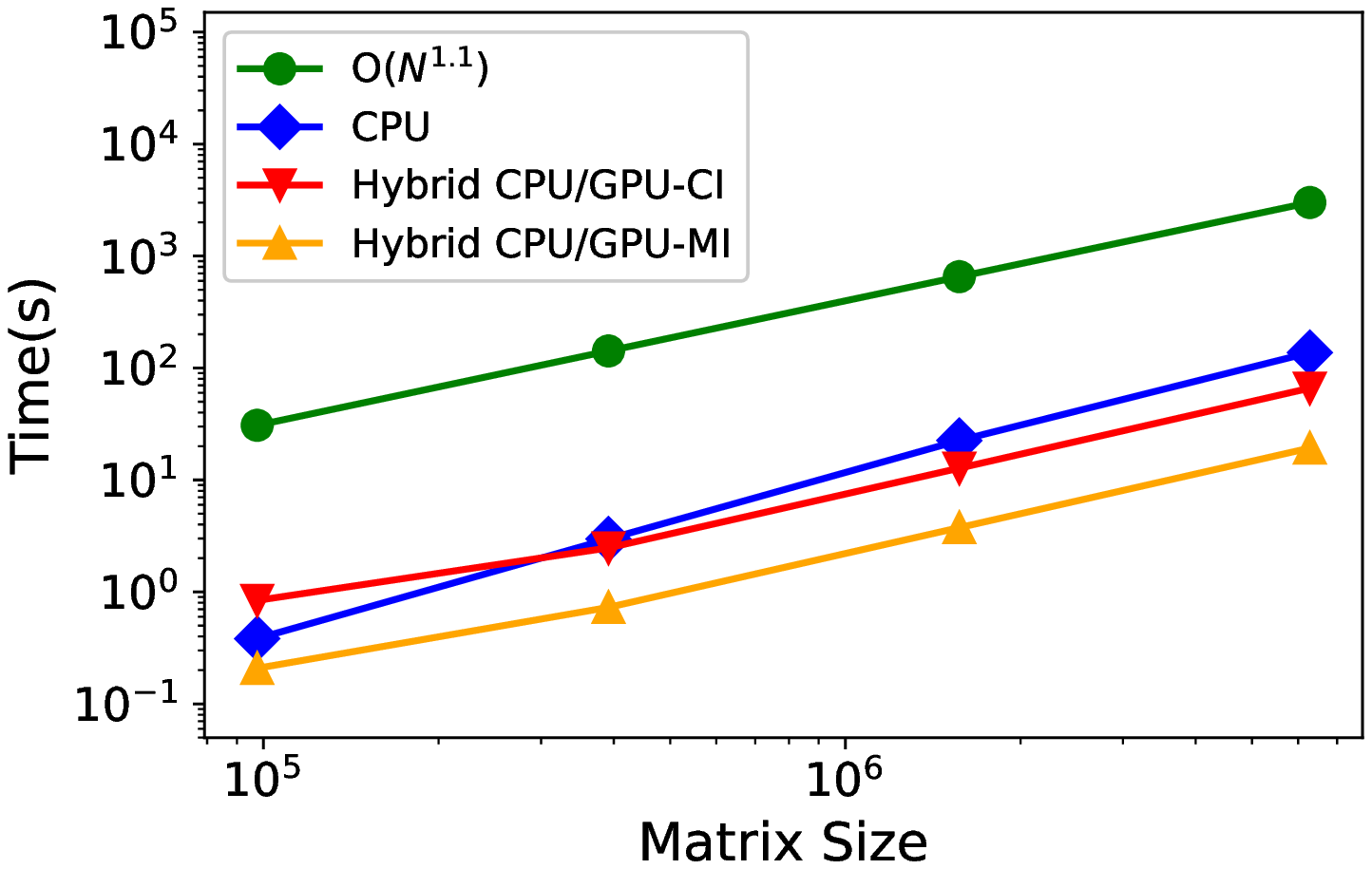}
		\vspace{-4mm}
		\caption{Complexity of AMG with P4 FE}
		\label{P2V4}
	\end{subfigure}
	\caption{Complexity of AMG solver for unsymmetric matrices}
	\label{P2}	
\end{figure}

Fig.~\ref{P1V} shows the time complexity of hybrid parallel implementations for symmetric matrices for different different sparsity patterns. 
The complexity of AMG in all test cases is found to be approximately $\mathcal{O}(N^{1.1})$, which  slightly deviates from the ideal complexity of $\mathcal{O}(N)$.
For smaller matrices Hybrid CPU/GPU-CI took more time than baseline implementation due to less computing workload and dominant data transfer. For matrix sizes larger than 1M, both hybrid implementations perform better than baseline. Among all, the performance of hybrid CPU/GPU-MI is better at the cost of additional GPU memory.
Next, Fig.~\ref{P2} shows the time complexity for unsymmetric matrices. We observe a same  complexity and similar performance as in the symmetric case even in the largest system of size 6M.

\begin{figure}
	\centering
	\begin{subfigure}{.5\textwidth}
		\centering
		\includegraphics[scale=0.4]{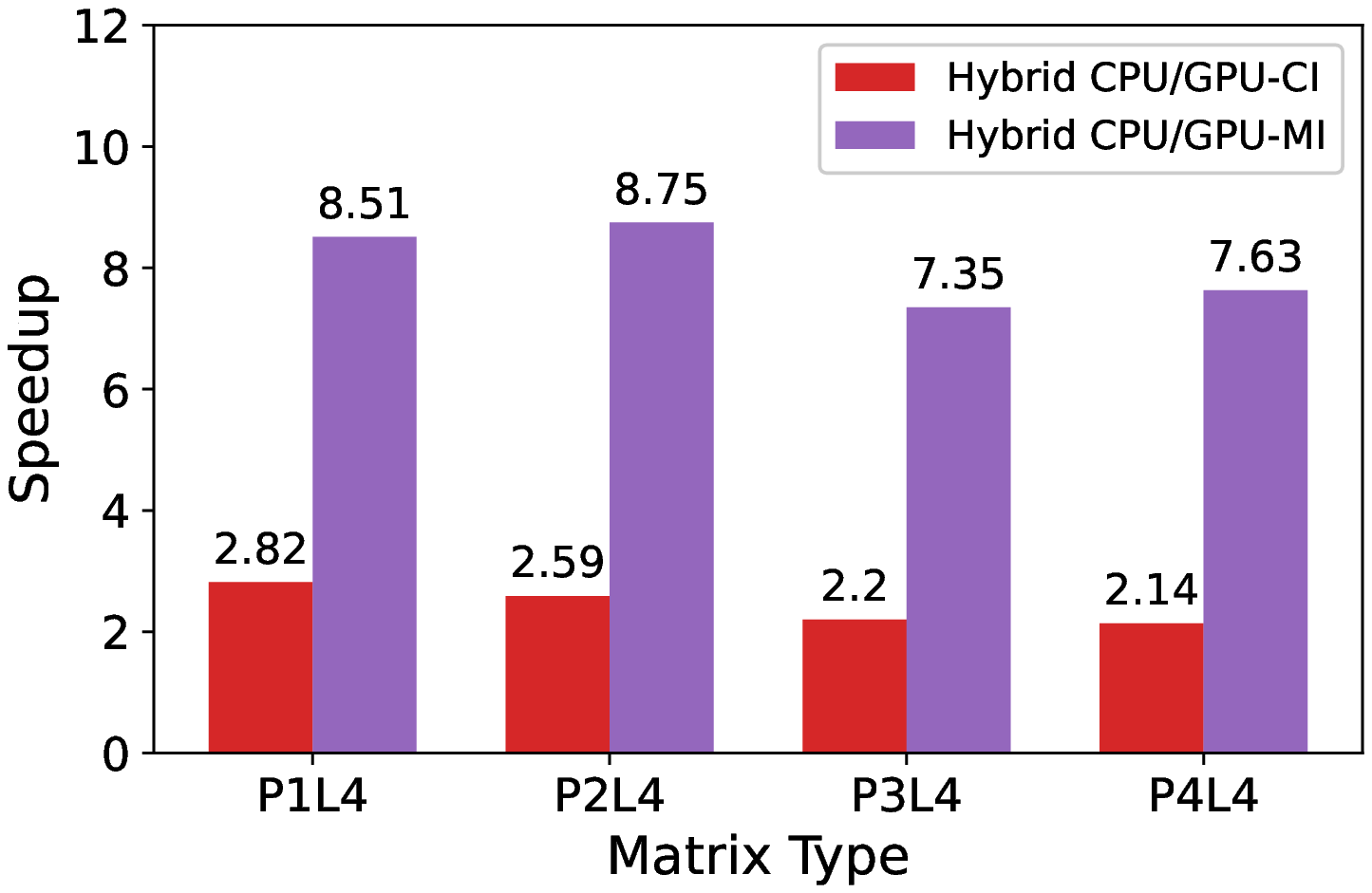}
		\vspace{-2mm}
		\caption{Speedup for symmetric matrices}
		\label{speedup1}
	\end{subfigure}%
	\begin{subfigure}{.5\textwidth}
		\centering
		\includegraphics[scale=0.4]{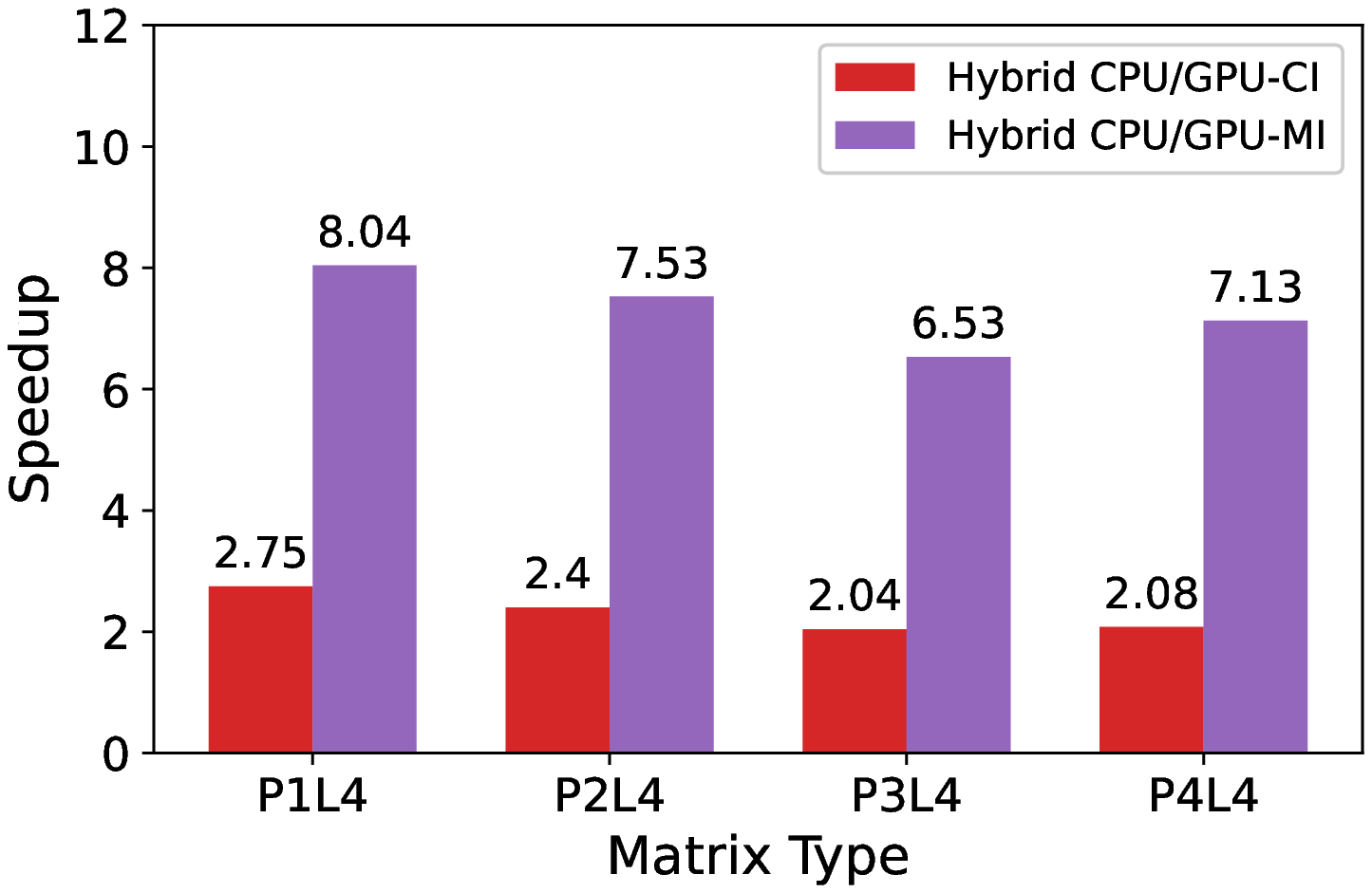}
		\vspace{-2mm}
		\caption{Speedup for unsymmetric matrices}
		\label{speedup2}
	\end{subfigure}
	\vspace{-3mm}
	\caption{ Speedups attained in hybrid implementations for matrices with different size and sparsity}
	\label{speedup}	
\end{figure}

Speedups obtained in each FE type with largest matrix size are depicted in Fig.~\ref{speedup}. Hybrid CPU/GPU-CI provides up to 2X reduction in computing time in all test cases compared to baseline implementation. Moreover, hybrid CPU/GPU-MI implementation provides up to 7X reduction in computing time at the cost of additional GPU memory usage.

\subsection{Performance of AMG-PCG}
\begin{figure}
	\centering
	\begin{subfigure}{.5\textwidth}
		\centering
		\includegraphics[scale=0.4]{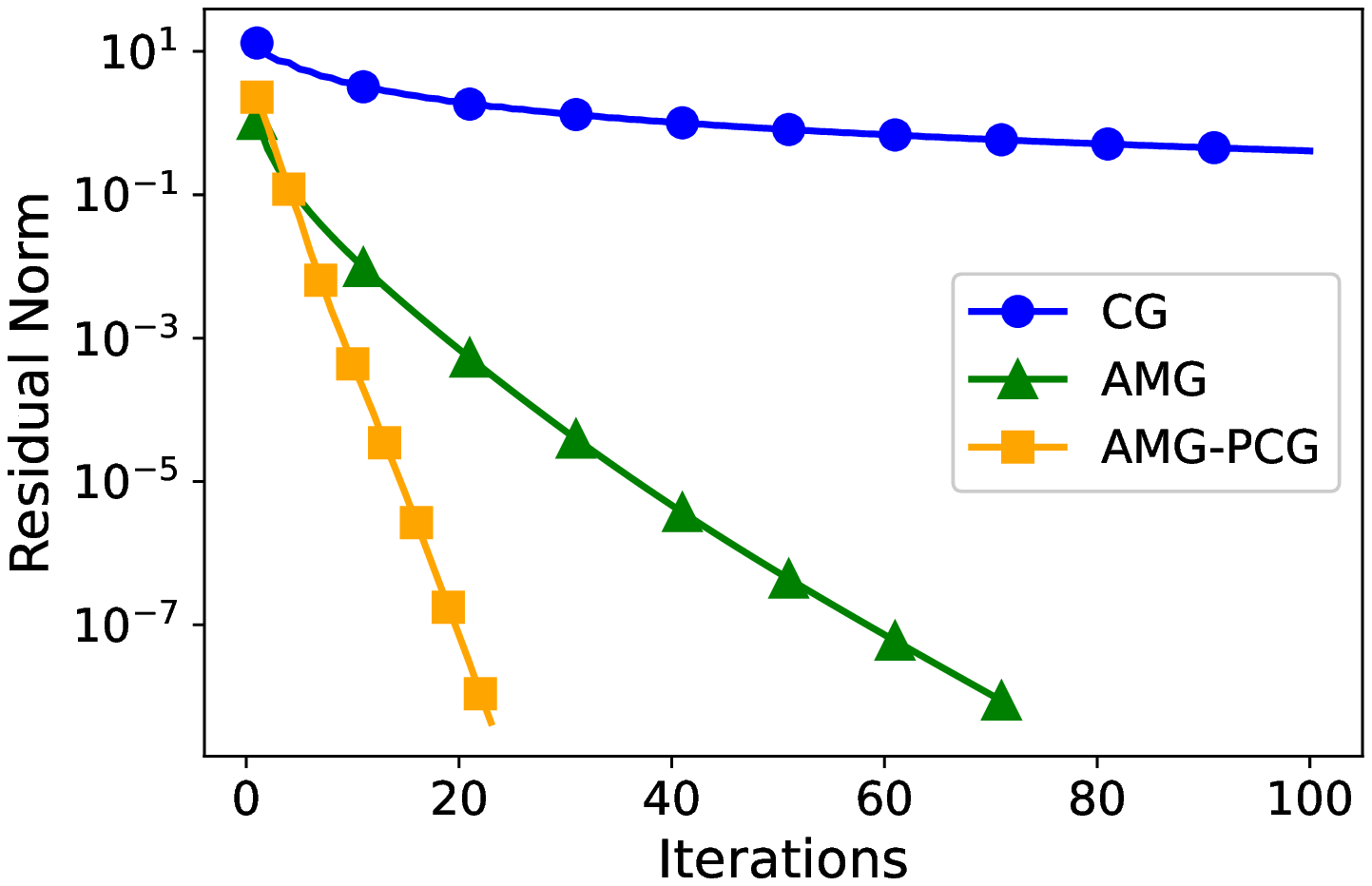}
		\vspace{-4mm}
		\caption{Comparison of convergence}
		\label{PCG_1}
	\end{subfigure}%
	\begin{subfigure}{.5\textwidth}
		\centering
		\includegraphics[scale=0.4]{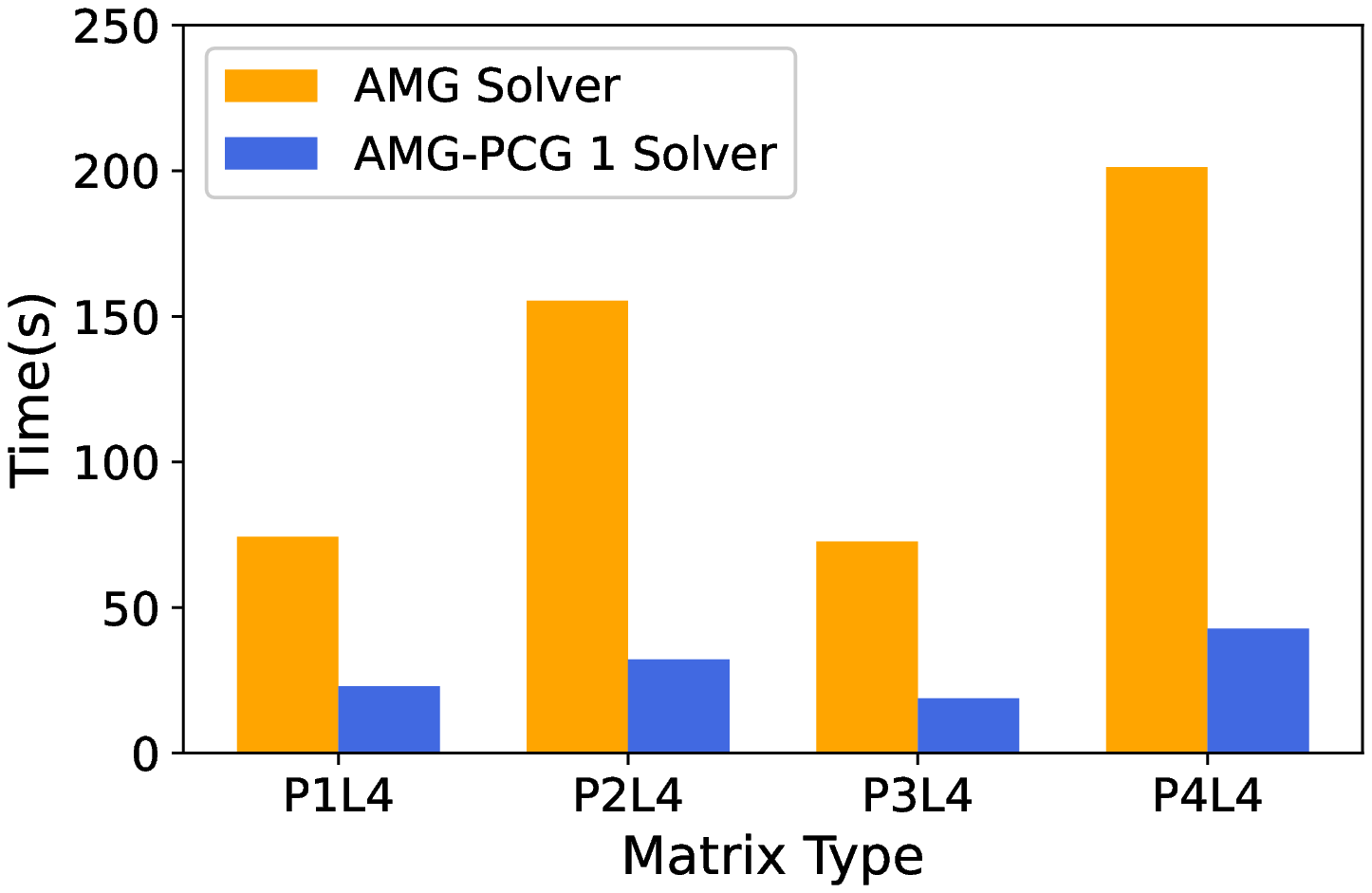}
		\vspace{-4mm}
		\caption{Time Comparison}
		\label{PCG_2}
	\end{subfigure}
	\caption{Comparison of AMG and AMG-PCG solvers}	
\end{figure}

Multigrid methods are found to perform better as preconditioner  to Krylov subspace methods. Since AMG does not require any other details of the   computational domain, it is often used as black box preconditioner to Krylov subspace methods. Fig~\ref{PCG_1} presents a comparison of Conjugate Gradient (CG) method, AMG as solver and AMG-PCG. Symmetric matrices of type P1L4, P2L4, P3L4 and P4L4 are considered to perform this study, where the computations are performed with baseline implementation.  
We can clearly see that AMG-PCG convergence much faster than  AMG   and CG solvers. Moreover, up to 6X reduction in computing time is obtained in AMG-PCG over AMG as solver as highlighted in Fig.~\ref{PCG_2}. Profiling of CPU implementation of AMG-PCG i.e. (AMG-PCG~1) reveals that it achieves  11.7\% of system peak performance with an arithmetic intensity of 0.13.

\begin{figure}
	\centering
	\begin{subfigure}{.5\textwidth}
		\centering
		\includegraphics[scale=0.4]{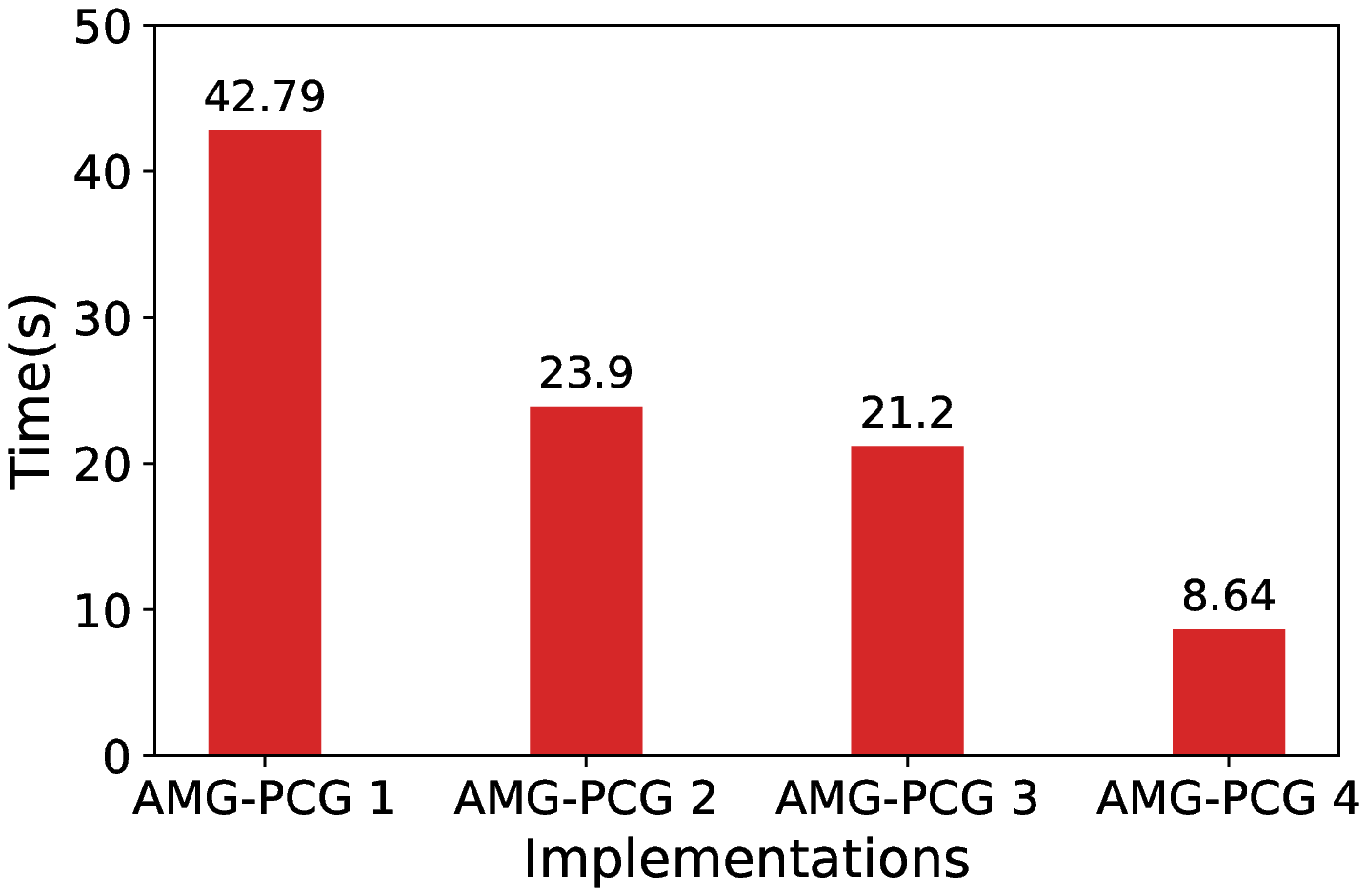}
		\vspace{-2mm}
		\caption{Comparison of Solution Time}
		\label{PCG_3}
	\end{subfigure}%
	\begin{subfigure}{.5\textwidth}
		\centering
		\includegraphics[scale=0.4]{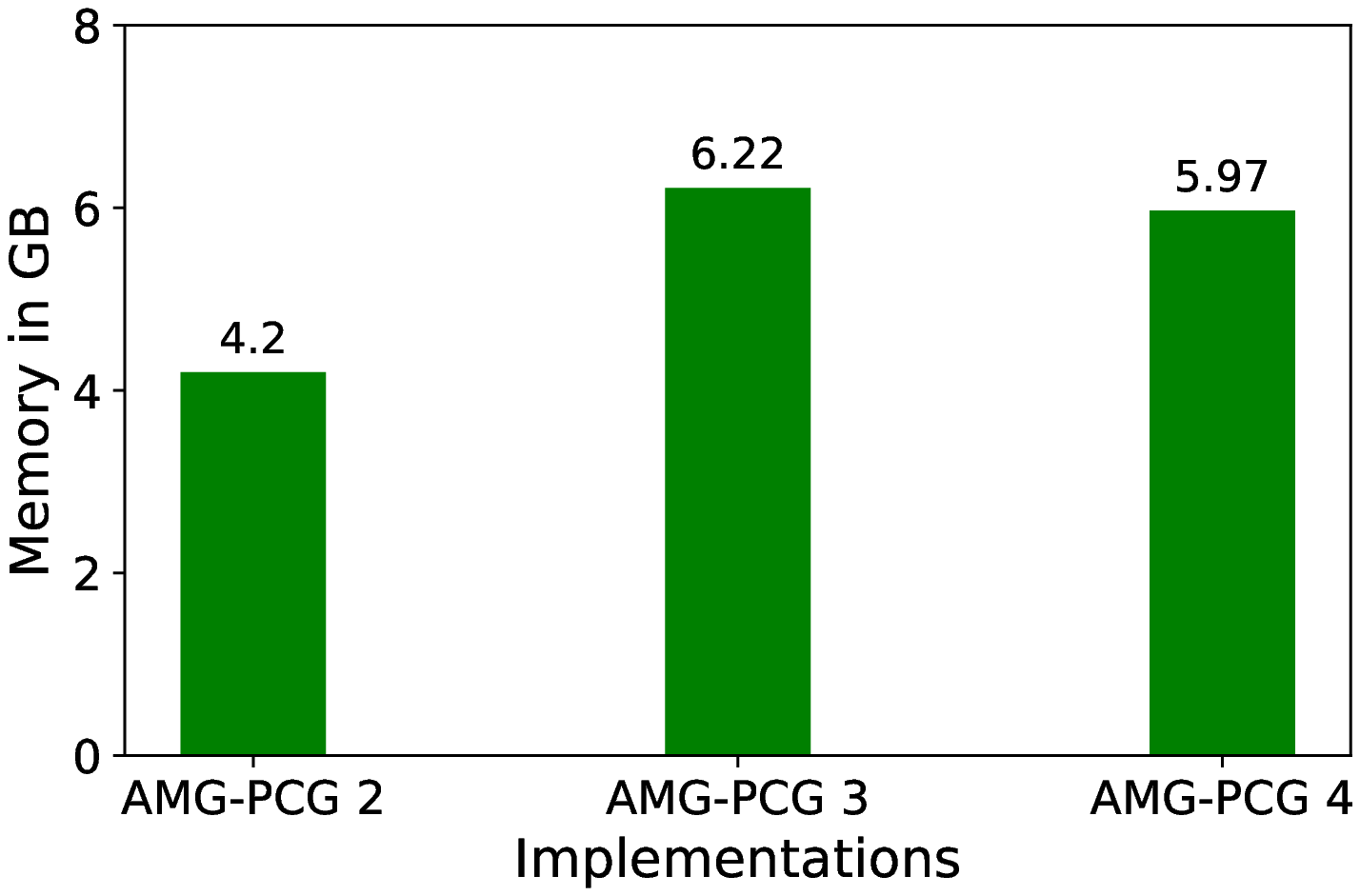}
		\vspace{-2mm}
		\caption{Comparison of GPU Memory requirements}
		\label{PCG_4}
	\end{subfigure}
	\caption{ Comparison of time and memory usage in AMG-PCG implementations}
\end{figure}

Furthermore, the four variants of AMG-PCG are evaluated on a symmetric matrix of type P4L4. The computing time and GPU memory requirements are compared in Fig.~\ref{PCG_3} and \ref{PCG_4}. Improvement of up to 2X reduction in computing time is obtained when AMG preconditioning is performed using hybrid CPU/GPU-CI approach in AMG-PCG 2, see Fig.~\ref{PCG_3}.
Improvements obtained are negligible in AMG-PCG 3 when CG steps are computed on GPU since the time taken by CG steps form a very small proportion of total computing time. However, it requires more GPU memory and overheads because of additional allocation of matrix and transfer to GPU to perform CG steps.
AMG-PCG 4 implementation takes less computing time among all four implementations but requires an additional GPU memory. Further, 3X reduction in computing time is obtained in AMG-PCG 4 compare to AMG-PCG 2 but at the cost of 33$\%$ more GPU memory than AMG-PCG 2 as shown in Fig.~\ref{PCG_4}.    
\subsection{Performance of AMG-PBiCG}
\begin{figure}
	\centering
	\begin{subfigure}{.5\textwidth}
		\centering
		\includegraphics[scale=0.4]{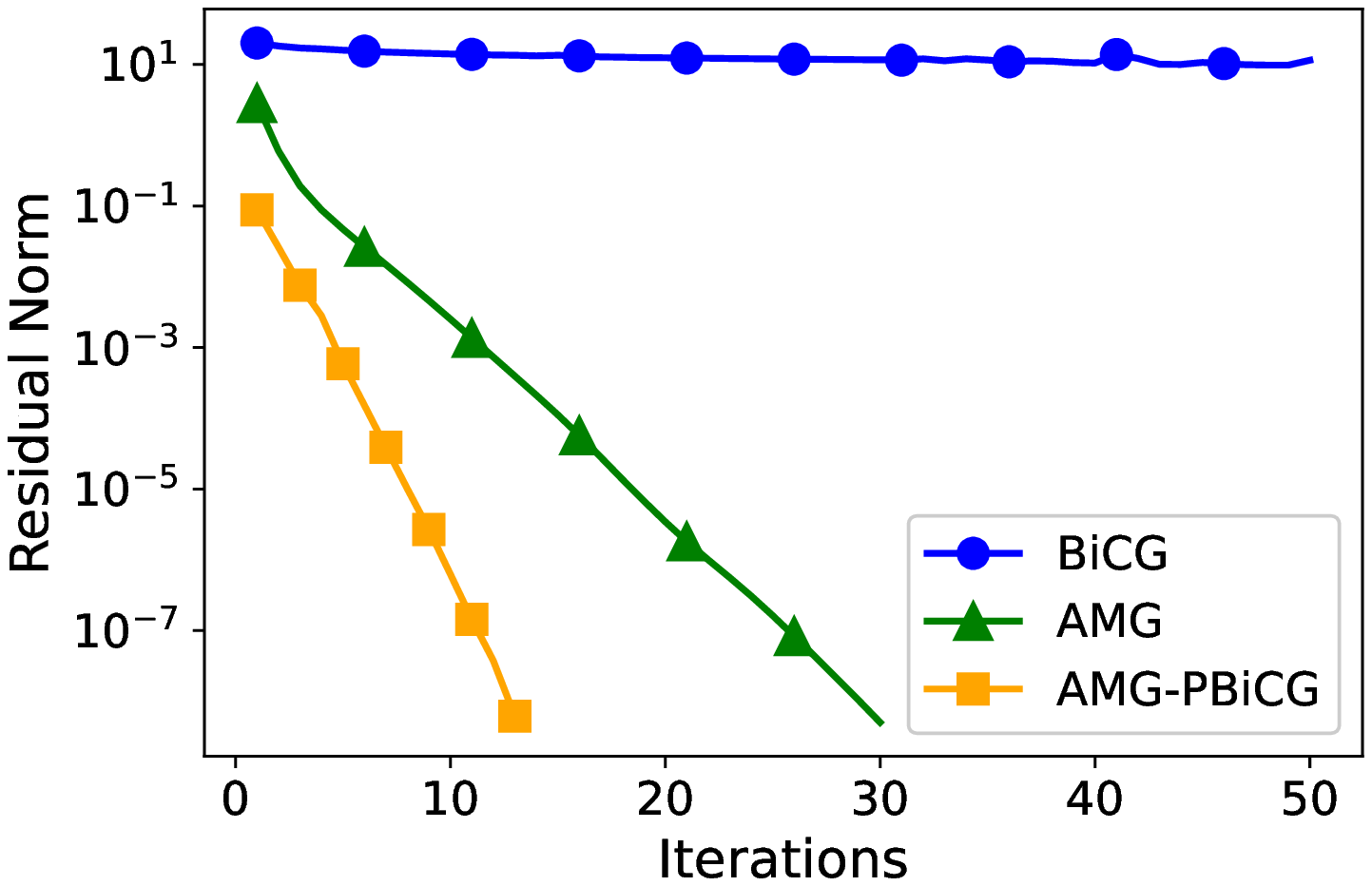}
		\vspace{-4mm}
		\caption{Comparison of convergence}
		\label{PBiCG_1}
	\end{subfigure}%
	\begin{subfigure}{.5\textwidth}
		\centering
		\includegraphics[scale=0.4]{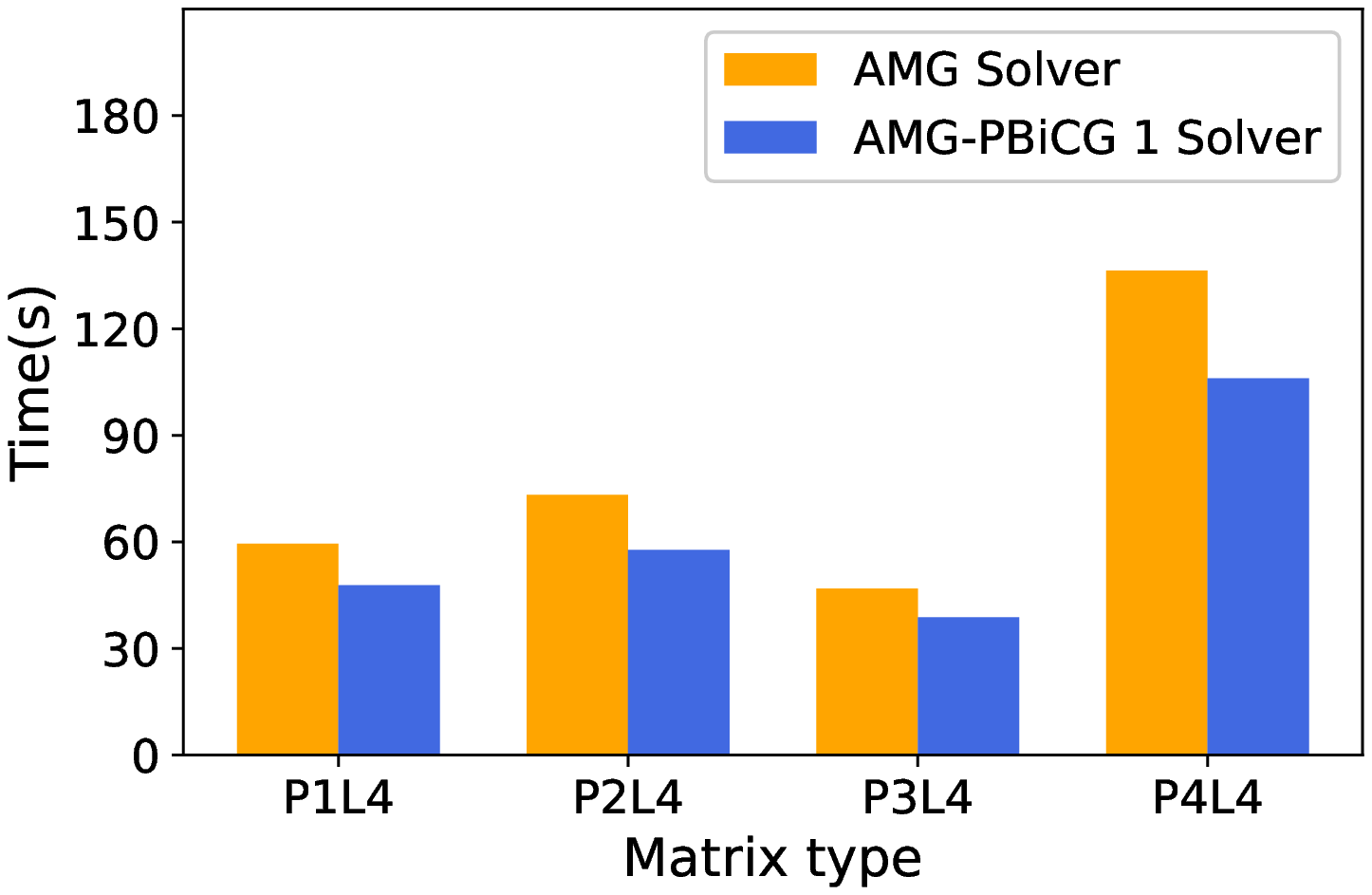}
		\vspace{-4mm}
		\caption{Time Comparison}
		\label{PBiCG_2}
	\end{subfigure}
	\vspace{-2mm}
	\caption{Comparison of AMG and AMG-PBiCG solvers}	
\end{figure}

Conjugate Gradient method requires system matrix to be symmetric and positive definite. Bi-Conjugate Gradient Stabilized (BiCG) method is a Krylov method which relaxes the symmetric constraint on the system matrix. Unsymmetric matrices of type P1L4, P2L4, P3L4 and P4L4 are considered to evaluate the performance of AMG-PBiCG.  
Convergence behaviour of BiCG solver, AMG as solver and AMG-PBiCG solver obtained using baseline implementation is shown in Fig.~\ref{PBiCG_1}.  Implementation of AMG as preconditioner to BiCG accelerates the convergence of BiCG method. Fig~\ref{PBiCG_2} shows the total computing time taken by AMG solver and AMG-PBiCG solver. We can observe up to 20\% reduction in computing time in AMG-PBiCG over AMG  solver.   

\begin{figure}
	\centering
	\begin{subfigure}{.5\textwidth}
		\centering
		\includegraphics[scale=0.4]{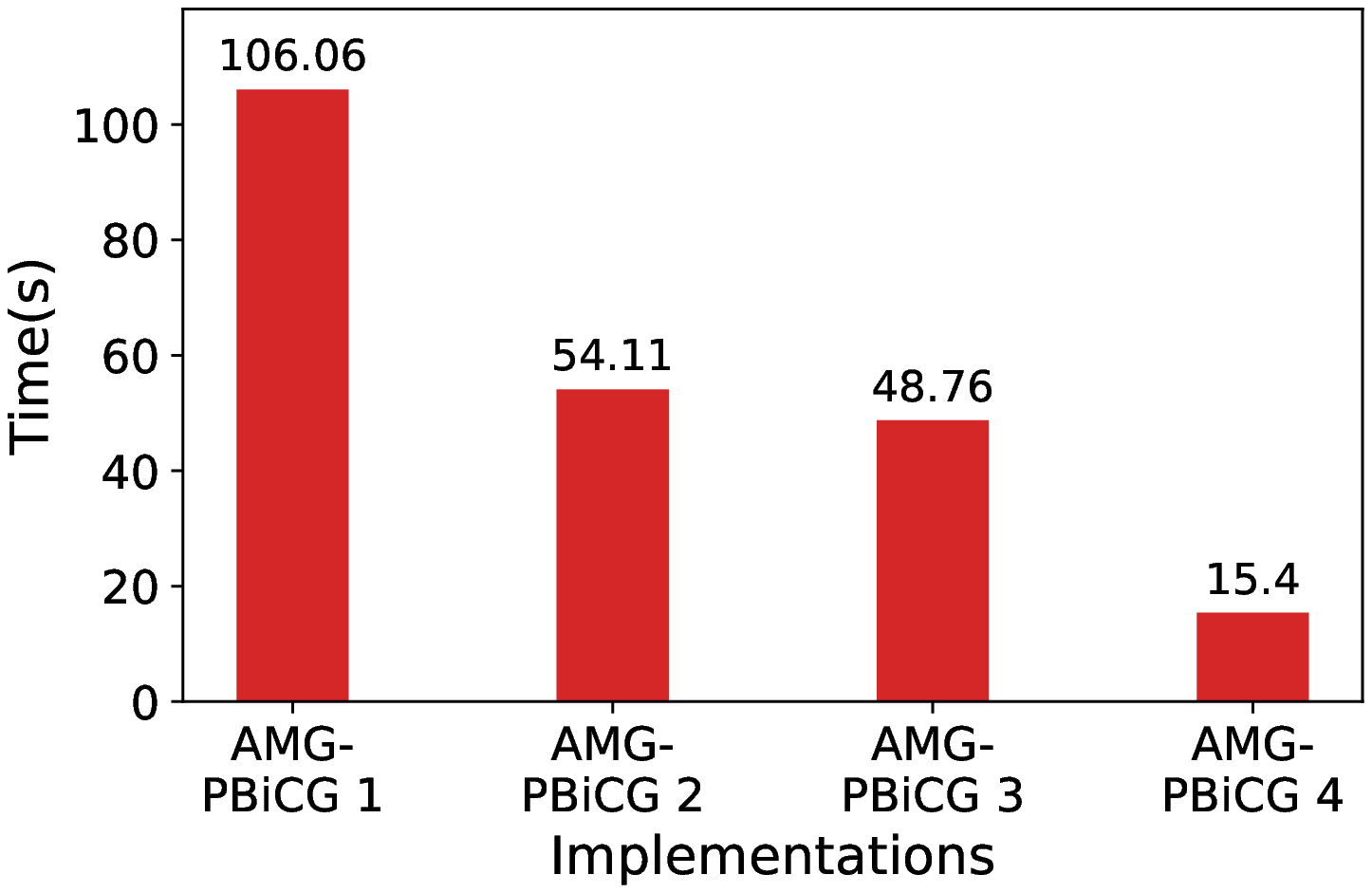}
		\vspace{-2mm}
		\caption{Comparison of Solution Time}
		\label{PBiCG_3}
	\end{subfigure}%
	\begin{subfigure}{.5\textwidth}
		\centering
		\includegraphics[scale=0.4]{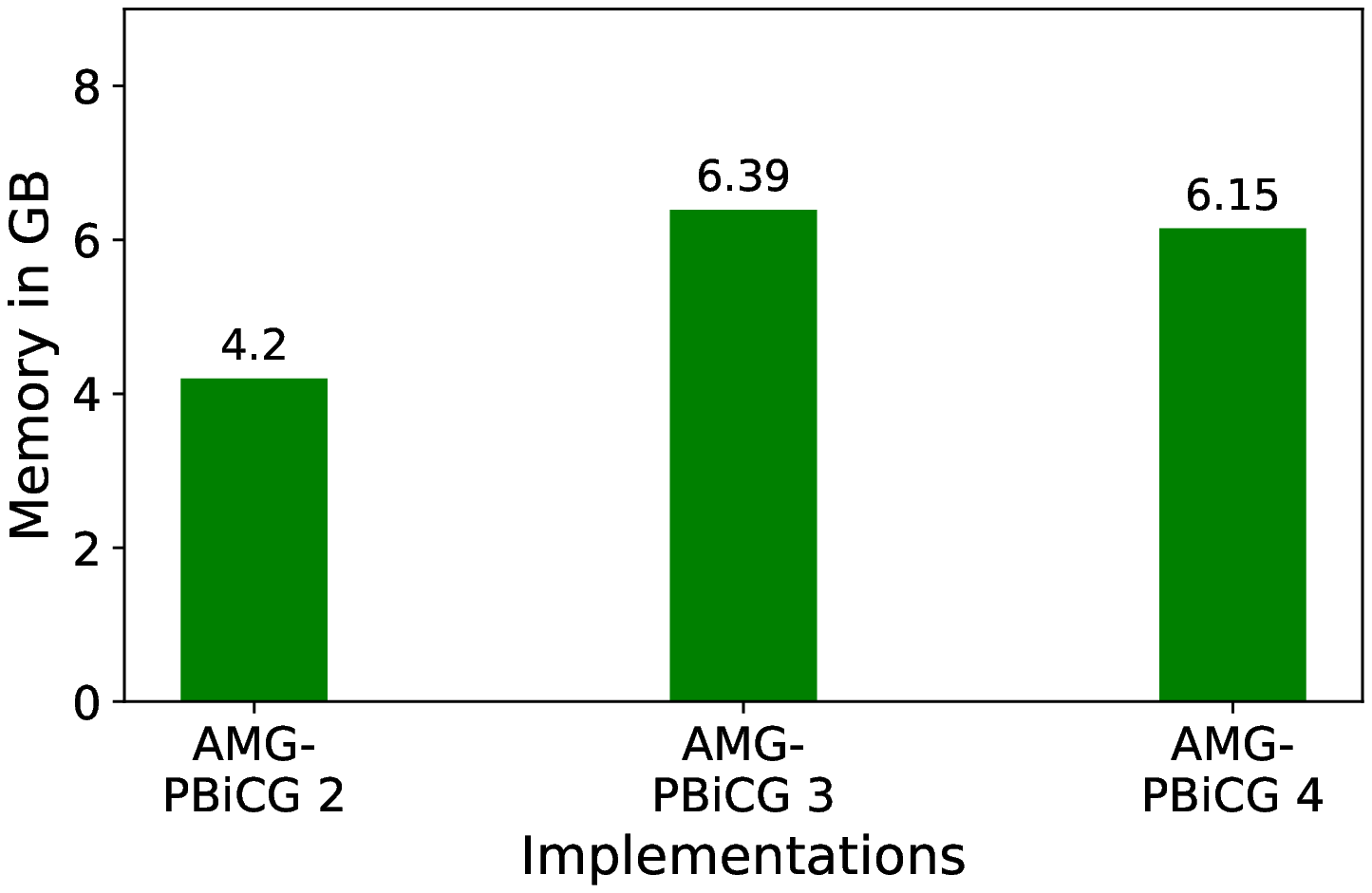}
		\vspace{-2mm}
		\caption{Comparison of GPU Memory requirements}
		\label{PBiCG_4}
	\end{subfigure}
	\caption{ Comparison of time and memory usage of AMG-PBiCG implementations}
	\label{PBiCG2}	
\end{figure}

The total computing time taken by four implementations of AMG-PBiCG show a similar trend observed in AMG-PCG implementations, see Fig.~\ref{PBiCG2}. Here, all four implementations are evaluated on an unsymmetric matrix of type P4L4. Further, 2X reduction in computing time is obtained when AMG preconditioning  is computed on GPU using hybrid CPU/GPU-CI algorithm. Profiling results reveal that a majority of computation time is taken by AMG preconditioning, whereas BiCG steps require relatively less time. Hence, performing this less compute intensive calculations on GPU (AMG-PBiCG~3) does not provide much improvement over AMG-PBiCG~2 implementation but result in additional GPU memory requirements due to memory allocations needed for BiCG steps as shown in Fig.~\ref{PBiCG_3} and \ref{PBiCG_4}. Further, AMG-PBiCG~4 gives upto 3X improvement over AMG-PBiCG~2 in terms of reduction in computation time but occupy additional amount of GPU memory.     

\subsection{Comparison of Hybrid CPU-GPU AMG with AMGX}
GPU-only implementation, AMGX is one of the modern AMG packages specifically designed to exploit NVIDIA GPU architectures. Such packages provide better performance gain than CPU-based AMG packages. 
In this  section, the performance of the proposed hybrid implementations of AMG, $i.e.$ Hybrid CPU/GPU-CI and  Hybrid CPU/GPU-MI are compared  with the GPU-only AMG package AMGX.

\begin{figure}
	\centering
	\begin{subfigure}{.5\textwidth}
		\centering
		\includegraphics[scale=0.4]{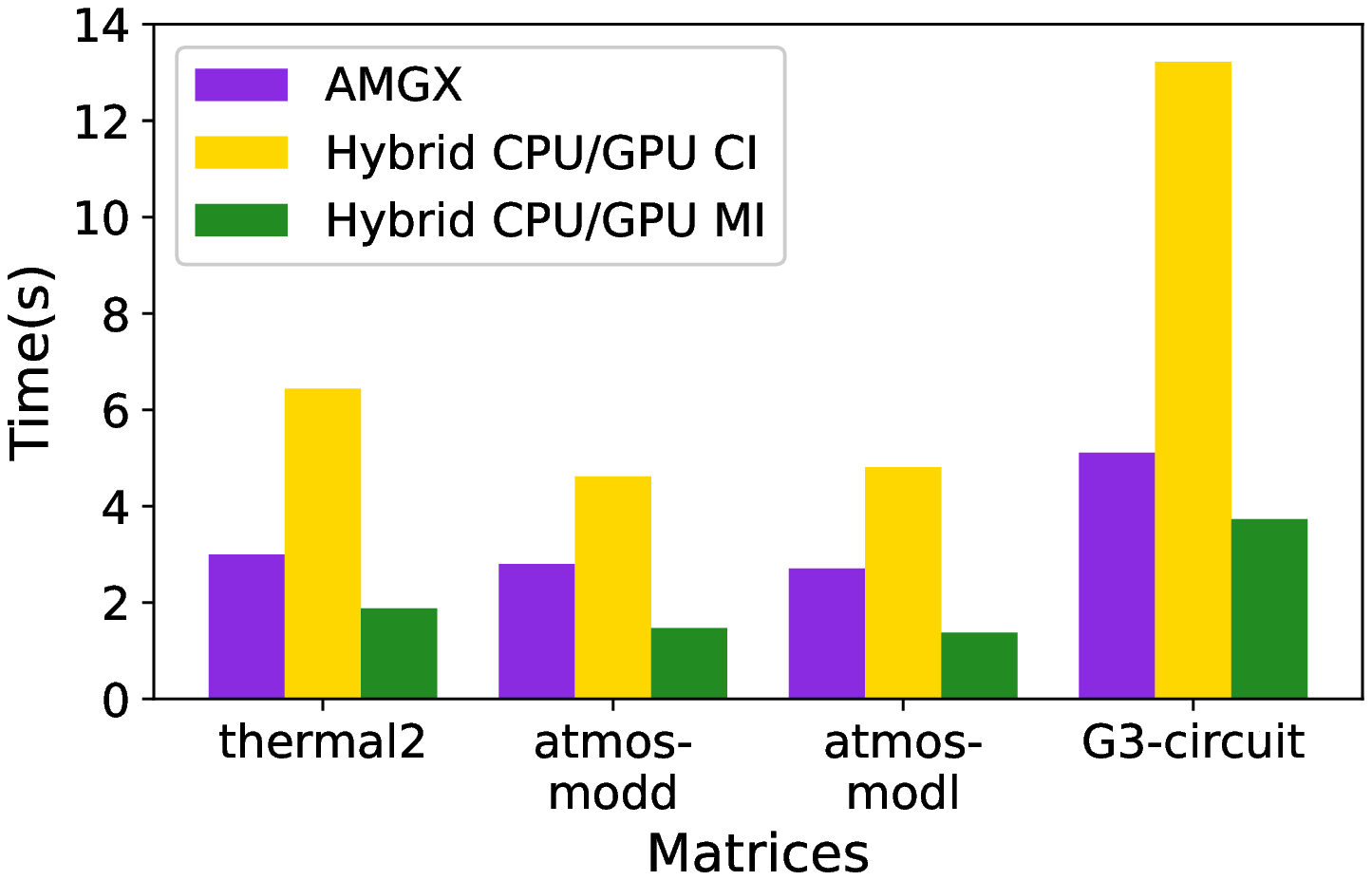}
		\vspace{-2mm}
		\caption{Comparison of Solution Time}
		\label{amgx_1}
	\end{subfigure}%
	\begin{subfigure}{.5\textwidth}
		\centering
		\includegraphics[scale=0.4]{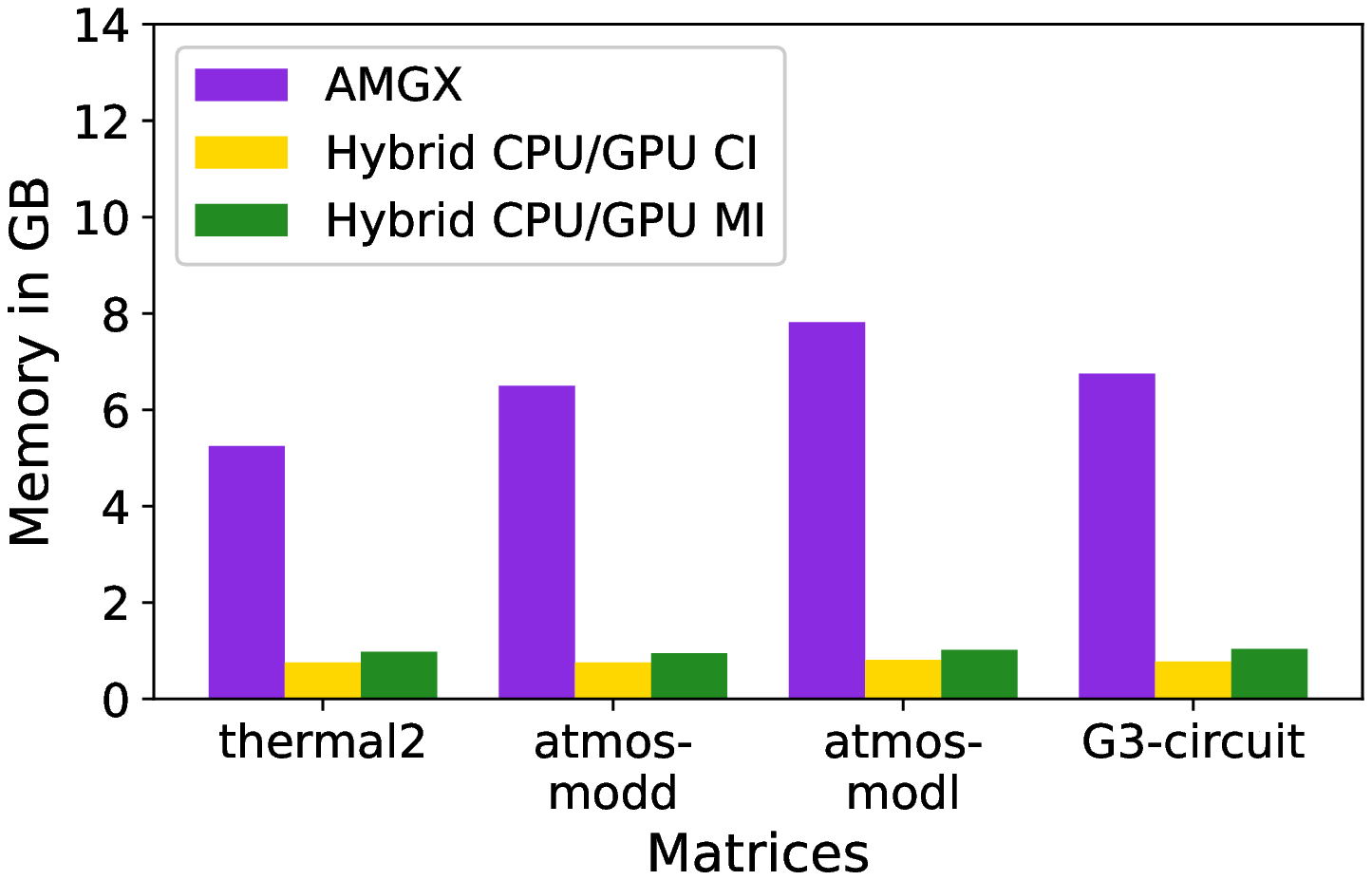}
		\vspace{-2mm}
		\caption{Comparison of GPU Memory requirements}
		\label{amgx_2}
	\end{subfigure}
	\caption{ Comparison of AMG and AMGX solvers for Sparse Suite collection matrices}
	\label{AMG_1}	
\end{figure}

\begin{figure}
	\centering
	\begin{subfigure}{.5\textwidth}
		\centering
		\includegraphics[scale=0.4]{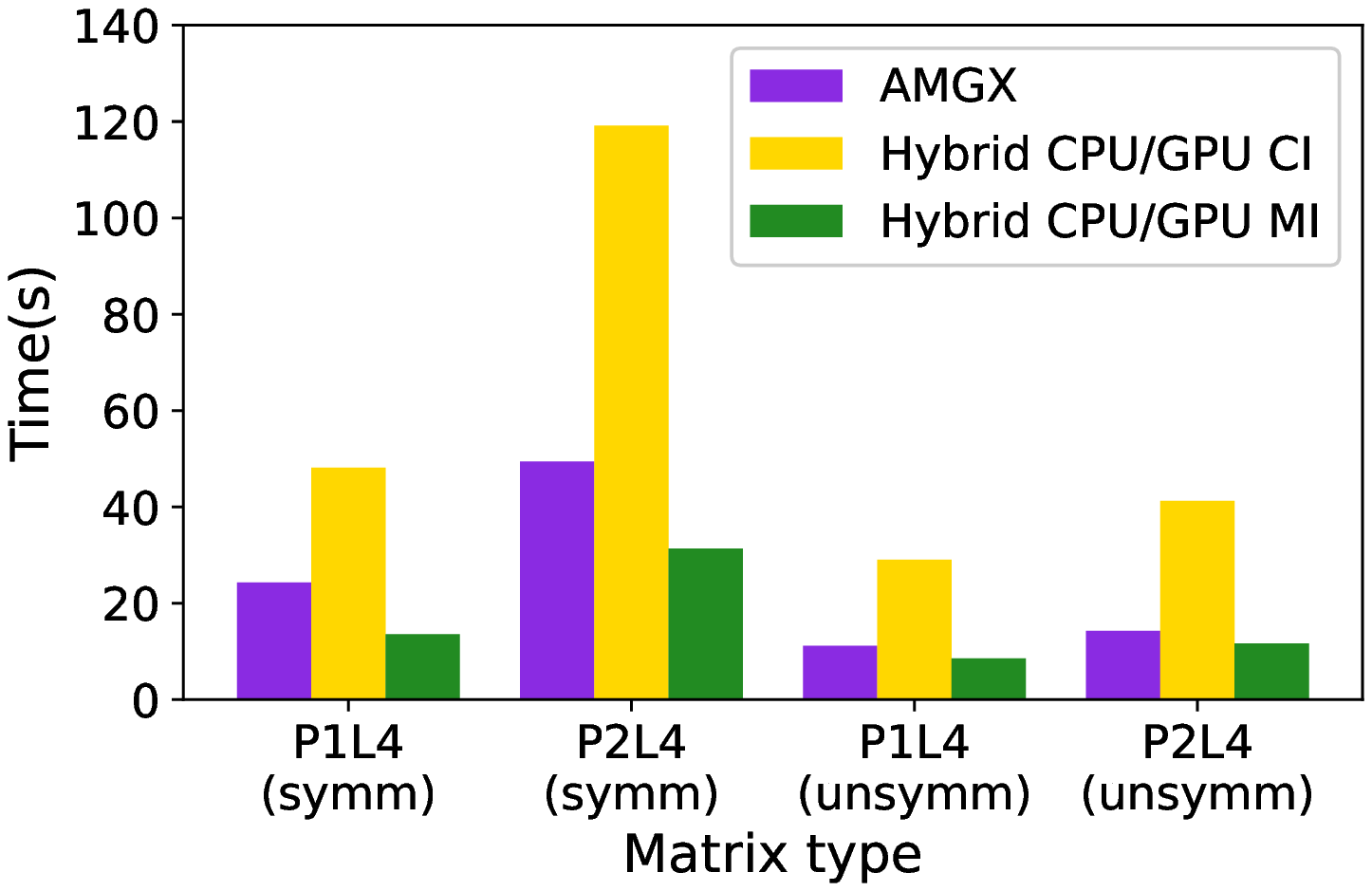}
		\vspace{-2mm}
		\caption{Comparison of Solution Time}
		\label{amgx_3}
	\end{subfigure}%
	\begin{subfigure}{.5\textwidth}
		\centering
		\includegraphics[scale=0.4]{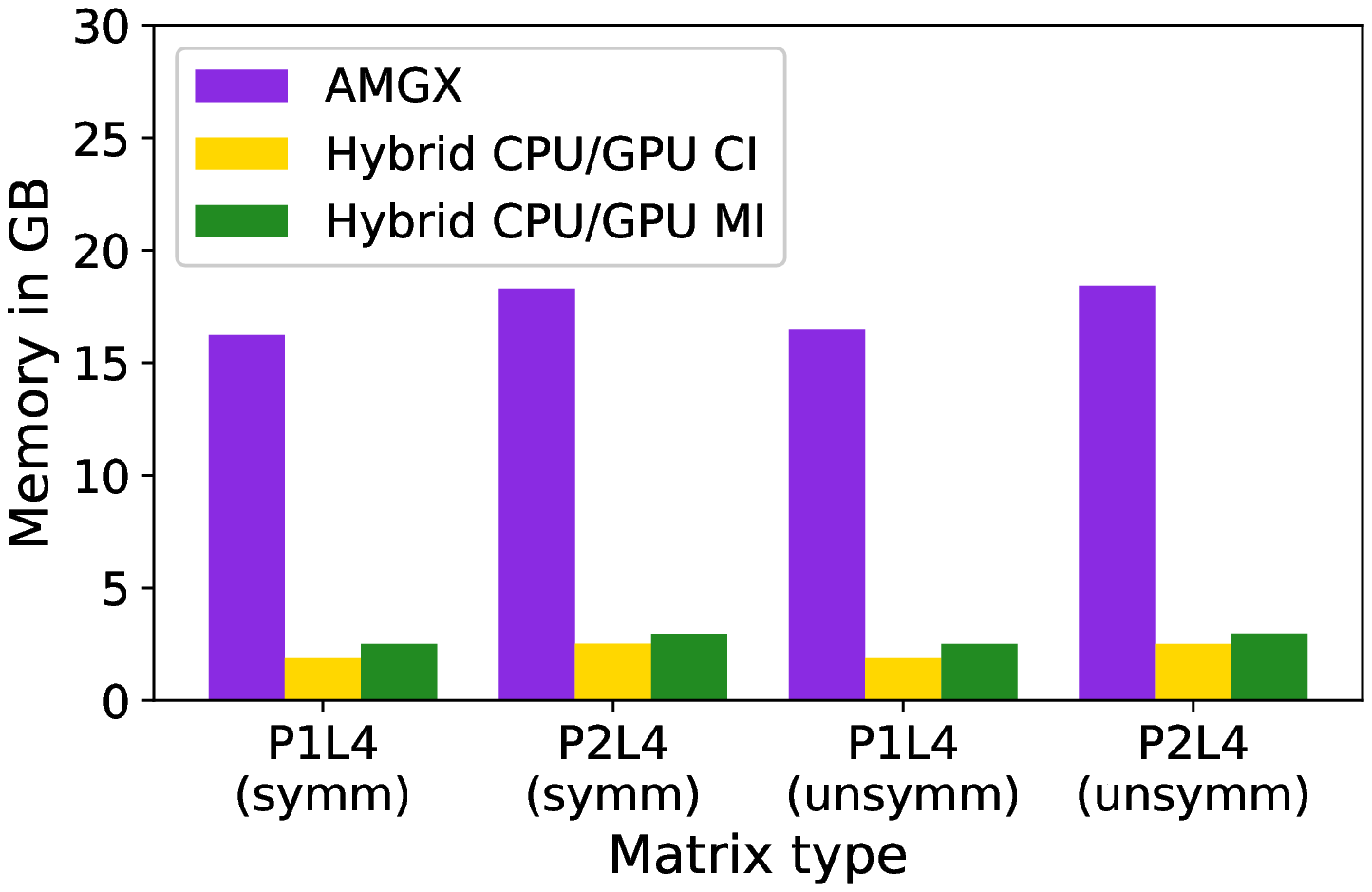}
		\vspace{-2mm}
		\caption{Comparison of GPU Memory requirements}
		\label{amgx_4}
	\end{subfigure}
	\caption{ Comparison of AMG as solver with AMGX}
	\label{AMG_2}	
\end{figure}

Matrices obtained from Sparse Suite Collection listed in Table~\ref{tab_4} are used in this comparative study. We first compare the hybrid CPU-GPU AMG as a solver with AMGX.  Fig.~\ref{amgx_1} compares the computing time of Hybrid AMG implementations with GPU-only implementations. Hybrid CPU/GPU-CI takes 2X more time to solve the linear systems but require only one-seventh of GPU device memory in comparison to GPU only implementations.  Contrarily, Hybrid CPU/GPU-MI implementation takes 20-30$\%$ less computing time than GPU-only implementations and more importantly  CPU/GPU-MI requires less GPU memory. Fig.~\ref{amgx_2} reveals that both hybrid implementations require significantly less GPU memory compared to GPU-only implementation. 
\begin{table}[tbh]
	\centering
	\caption{Test Matrices obtained from Sparse Suite Collection}\label{tab_4}
	\begin{tabular}{ccc}
		\hline
		\textbf{Matrix Name} & \textbf{Matrix Size} &  \textbf{Non zeros}\\ 
		\hline
		thermal2  &	1228045&  85803135\\ 
		atmosmodd &	1270432&  8814880\\ 
		atmosmodl &	1489752&  10319760\\ 
		G3-circuit&	1585478&  7660826\\ 
		\hline
	\end{tabular}
\end{table} 
More experiments are performed with symmetric and unsymmetric matrices of type P1L4, P2L4. Fig.~\ref{amgx_3} and Fig.~\ref{amgx_4} show the computing time and GPU memory requirements. As observed in the above comparison, both hybrid implementations require significantly less GPU memory compared to GPU-only implementation.  


\begin{figure}
	\centering
	\begin{subfigure}{.5\textwidth}
		\centering
		\includegraphics[scale=0.4]{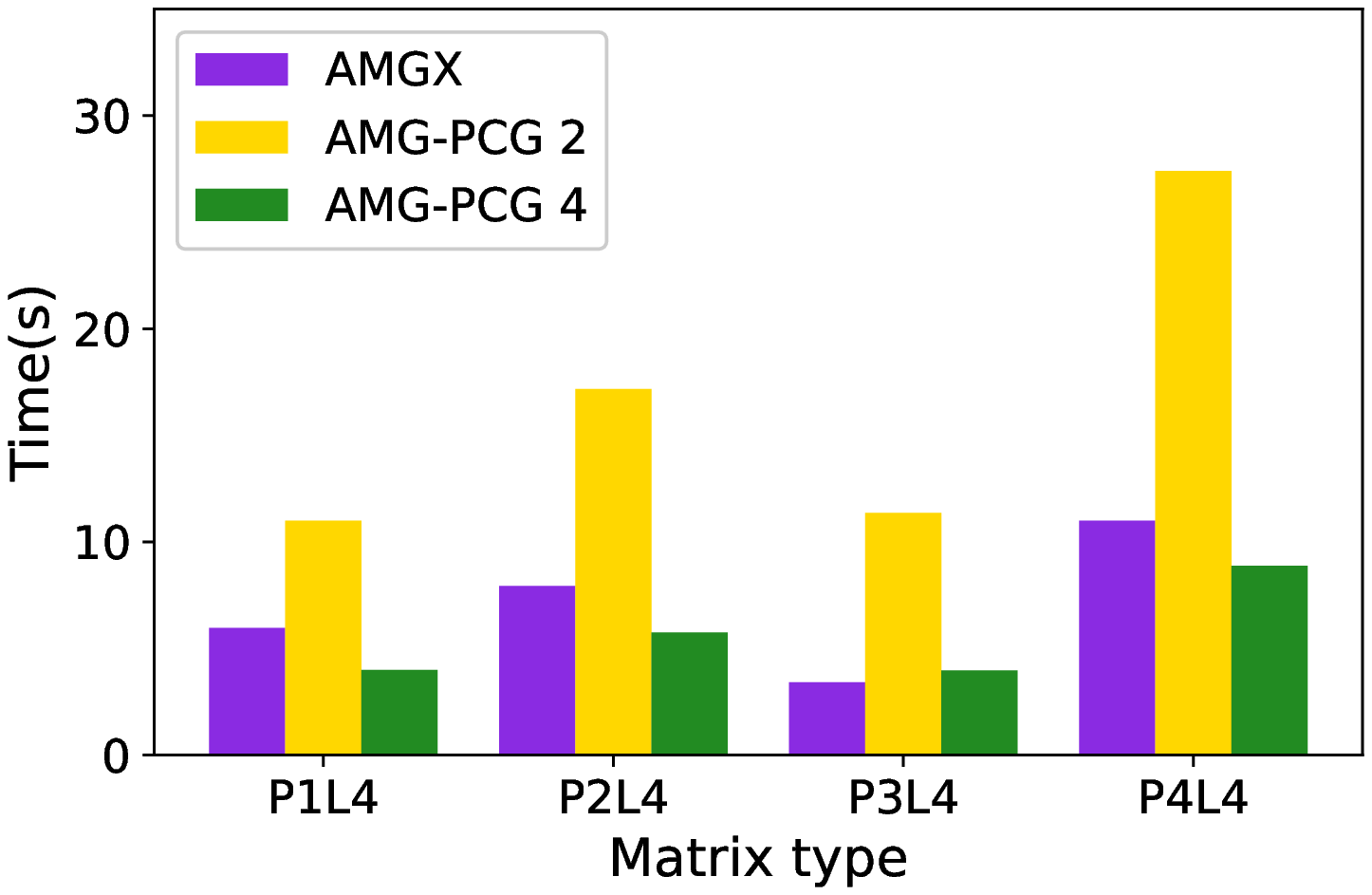}
		\vspace{-2mm}
		\caption{Comparison of Solution Time}
		\label{amgx_5}
	\end{subfigure}%
	\begin{subfigure}{.5\textwidth}
		\centering
		\includegraphics[scale=0.4]{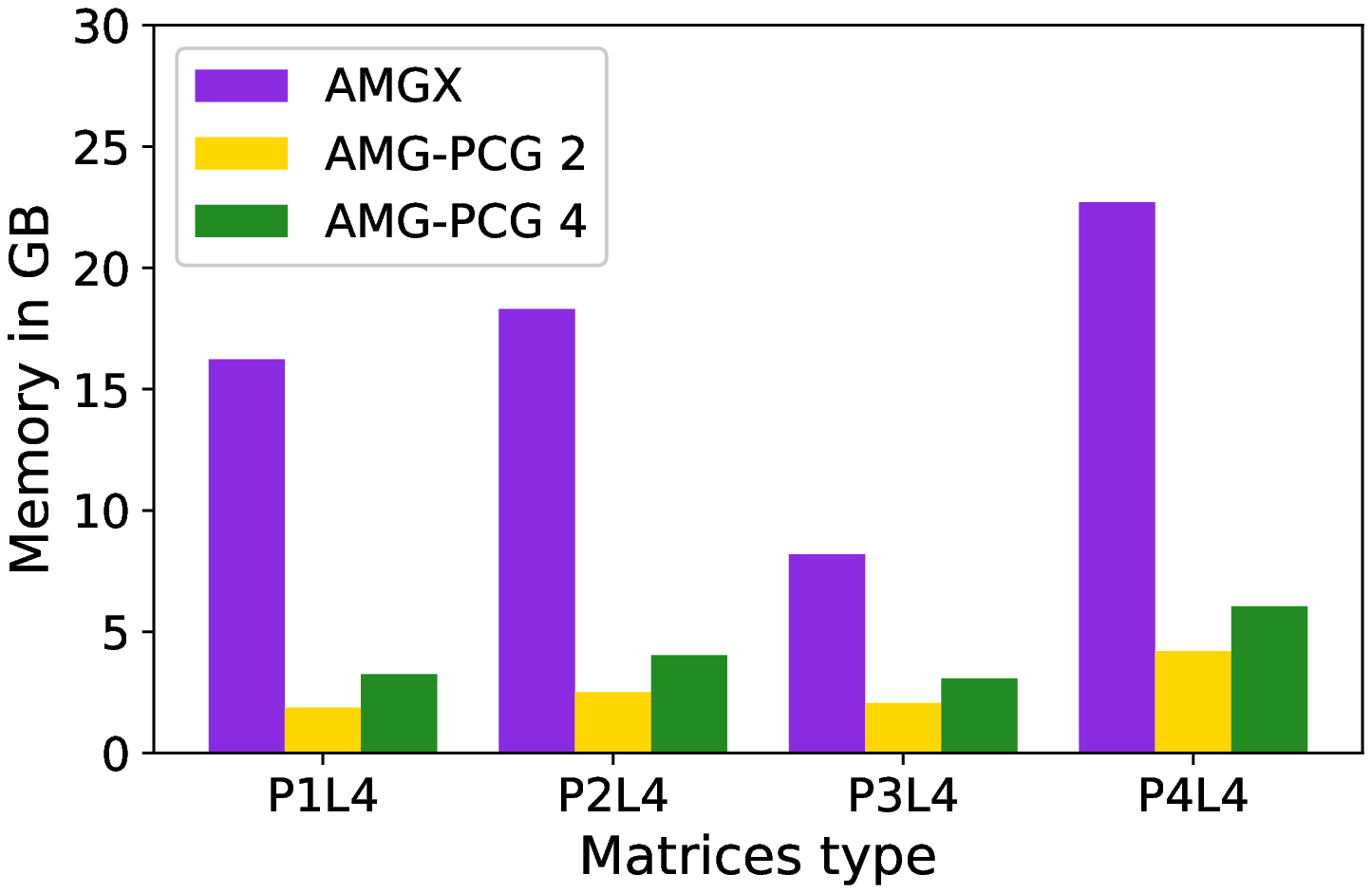}
		\vspace{-2mm}
		\caption{Comparison of GPU Memory requirements}
		\label{amgx_6}
	\end{subfigure}
	\caption{ Comparison of AMG-PCG implementations}
	\label{AMG_3}	
\end{figure}

\begin{figure}
	\centering
	\begin{subfigure}{.5\textwidth}
		\centering
		\includegraphics[scale=0.4]{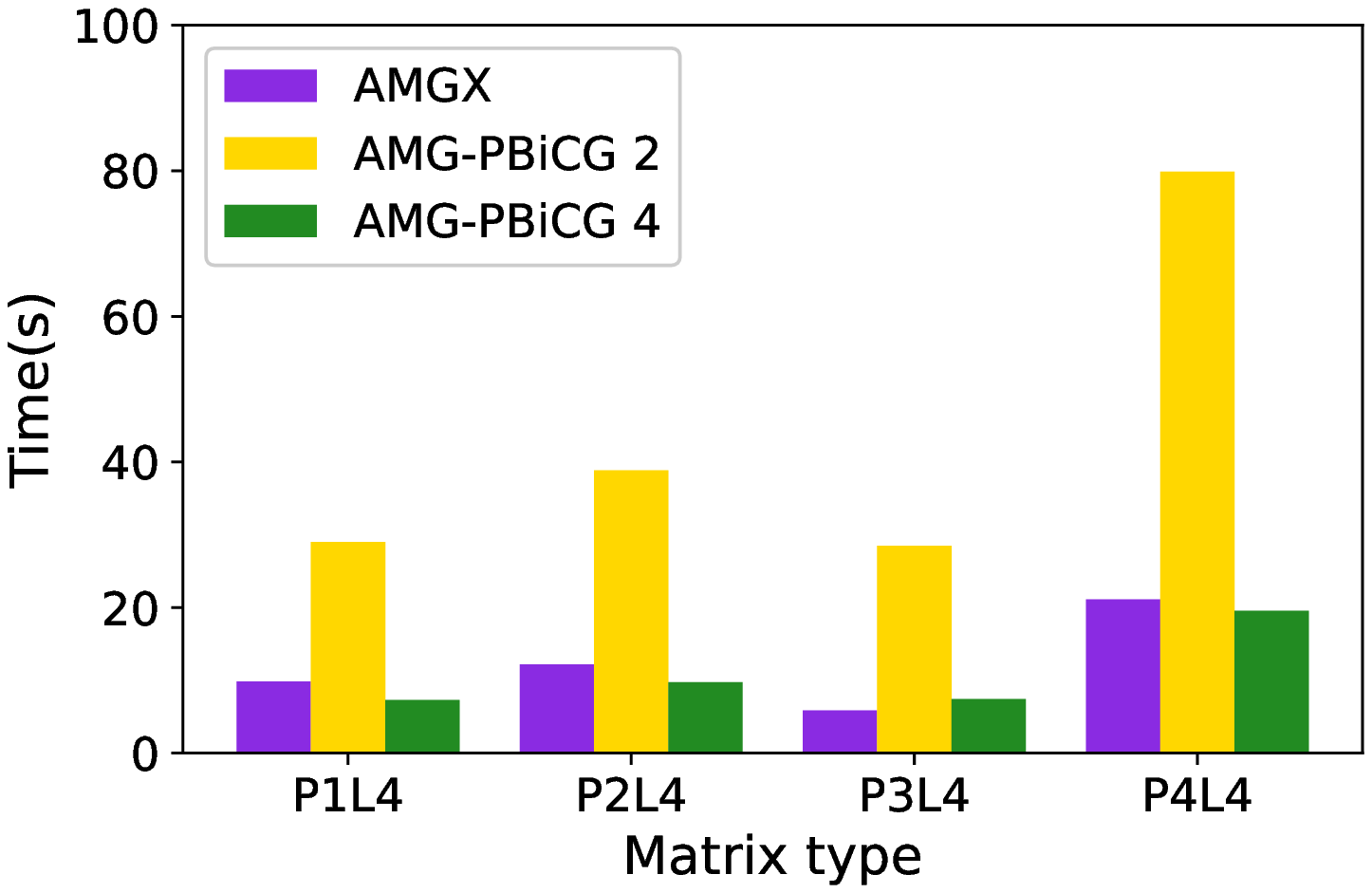}
		\vspace{-2mm}
		\caption{Comparison of Solution Time}
		\label{amgx_7}
	\end{subfigure}%
	\begin{subfigure}{.5\textwidth}
		\centering
		\includegraphics[scale=0.4]{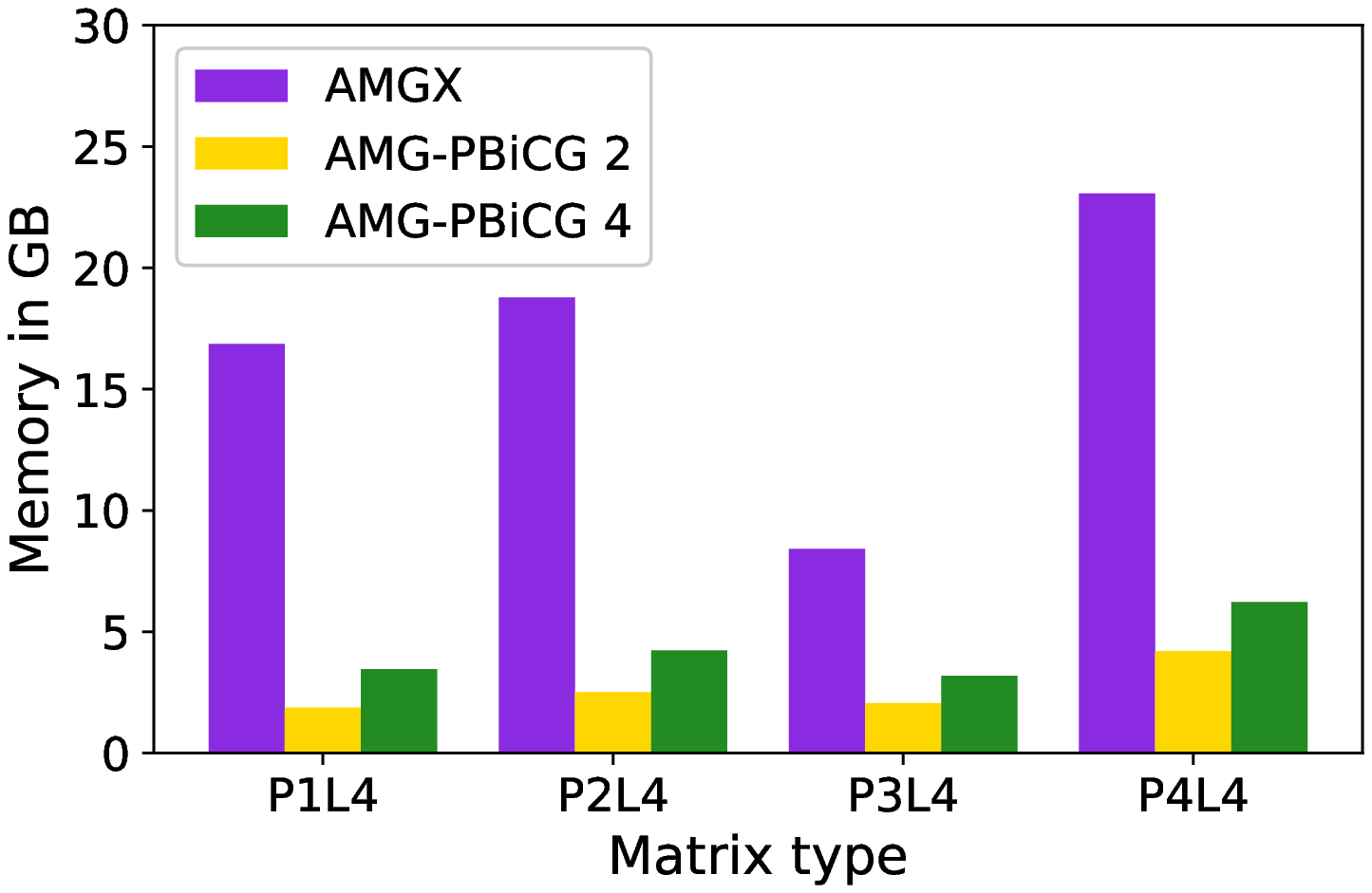}
		\vspace{-2mm}
		\caption{Comparison of GPU Memory requirements}
		\label{amgx_8}
	\end{subfigure}
	\caption{ Comparison of AMG-PBiCG implementations}
	\label{AMG_4}	
\end{figure}

We finally compare the performance of hybrid CPU-GPU implementations with AMGX as preconditioner to Krylov subspace solvers. In particular, AMG-PCG and AMG-PBiCG solver are compared. Symmetric and unsymmetric matrices of type P1L4, P2L4, P3L4 and P4L4 are used for AMG-PCG and AMG-PBiCG solvers, respectively.
Fig.~\ref{AMG_3} presents the computing time and GPU memory requirements for AMG-PCG~2, AMG-PCG~4 and the respective GPU-only implementation. Fig.~\ref{AMG_4} shows the computing time and GPU memory requirements for AMG-PBiCG~2, AMG-PBiCG~4 and the respective GPU-only implementation.
Hybrid AMG-PCG~2 and AMG-PBiCG~2 frameworks, which are based on CPU/GPU-CI implementations,  can be used in scenarios where lower GPU memory is available and when applications operating at the limits of available GPU memory. AMG-PCG~4 and AMG-PBiCG~4 frameworks, which are based on CPU/GPU-MI enable us to achieve same performance as compared to GPU-only implementation but with a significantly low GPU memory.

Overall, hybrid implementations enable to optimally utilize the available system resources without compromising the performance of the solvers. 
In large scale applications, both CPU and GPU resources can be used together in the proposed hybrid framework to cater the need of high computational resource associated with the application. 


\section{Conclusion}
\label{sec_conc}
Hybrid CPU-GPU parallel implementations of AMG solver suitable for modern day accelerator equipped computing systems are presented in this work. 
Further, two variants of pairwise aggregation based coarsening are presented
These implementations are designed to selectively perform certain computations on CPU consequently reducing the GPU memory requirements without compromising performance of the solver. 
For considered model problems, we have attained 7-8X speedup over CPU implementation with 16 OpenMP threads.  
Further GPU memory usage in hybrid implementations is one-seventh of GPU-only implementation and thus enables to solve large scale problems on the same device.
The proposed hybrid AMG framework is also used as preconditioner to Conjugate Gradient and Biconjugate Gradient iterative methods. The proposed library can be used as a standalone solver or can be integrated with existing PDE software packages. These solvers are integrated with our in-house finite element package, ParMooN. Our further work is focused towards designing such strongly coupled CPU-GPU implementations for distributed systems. 

The SParSH-AMG library presented in this paper can be downloaded at \\
 \href{https://github.com/parmoon/SParSH-AMG}{https://github.com/parmoon/SParSH-AMG}





\bibliographystyle{siamplain}
\bibliography{references}
\end{document}